\documentclass[letterpaper,aps,prc,superscriptaddress,nofootinbib,showpacs,floatfix]{revtex4-1}% PRC format
\usepackage[utf8]{inputenc}
\usepackage{amsmath}
\usepackage{amsfonts}
\usepackage{amssymb}
\usepackage{amsthm}
\usepackage{bm}
\usepackage{hyperref}
\usepackage[english]{babel}
\usepackage{fontenc}
\usepackage{graphicx}
\usepackage[margin=1in]{geometry}
\usepackage{enumerate}
\usepackage{textcomp}
\usepackage[toc,page]{appendix}
\usepackage{slashed}
\usepackage{color}

\usepackage{setspace}
\hypersetup{colorlinks=true, linkcolor=blue, citecolor=blue, urlcolor=blue}
{\large }

\begin{document}
\title{Mixed Species Charge and Baryon Balance Functions Studies with PYTHIA}

\author{ Claude~Pruneau }
\email{claude.pruneau@wayne.edu}
\affiliation{Department of Physics and Astronomy, Wayne State University, Detroit, 48201, USA}
\author{ Sumit~Basu }
\email{sumit.basu@cern.ch}
\affiliation{Lund University, Department of Physics, Division of Particle Physics, Box 118, SE-221 00, Lund, Sweden}
\author{ Victor~Gonzalez }
\email{victor.gonzalez@cern.ch}
\affiliation{Department of Physics and Astronomy, Wayne State University, Detroit, 48201, USA}
\author{ Brian~Hanley }
\email{bghanley@wayne.edu}
\affiliation{Department of Physics and Astronomy, Wayne State University, Detroit, 48201, USA}
\author{Ana Marin} 
\email{a.marin@gsi.de}
\affiliation{GSI Helmholtzzentrum f\"ur Schwerionenforschung GmbH, Research Division and ExtreMe Matter Institute EMMI, Darmstadt, Germany}
\author{ Alexandru F. Dobrin }
\email{alexandru.florin.dobrin@cern.ch}
\affiliation{Institute of Space Science -- INFLPR Subsidiary, Magurele, 077125, Romania}
\author{ Alexandru Manea }
\email{alexandru.manea@cern.ch}
\affiliation{Institute of Space Science -- INFLPR Subsidiary, Magurele, 077125, Romania}
\date{Nov 20, 2023}

\begin{abstract}
 Mixed species charge and baryon balance functions are computed based on proton--proton (pp) collisions simulated with the PYTHIA8 model. Simulations are performed with selected values of the collision energy $\sqrt{s}$ and the Monash tune and the Ropes and Shoving modes of PYTHIA8 to explore whether such measurements provide useful new information and constraints on mechanisms of particle production in pp collisions. Charge balance functions are studied based on mixed pairs of pions, kaons, and protons, whereas baryon balance functions are computed for mixed low mass strange and non-strange baryons. Both charge and baryon balance functions of mixed particle pairs feature shapes and amplitudes that sensitively depend on the particle considered owing largely to the particle production mechanisms implemented in PYTHIA. The evolution of balance functions integrals with the longitudinal width of the acceptance are presented and one finds that sums of such integrals for a given reference particle obey expected sum rules for both charge and baryon balance functions. Additionally, both types of balance functions are found to evolve in shape and amplitude with increasing collision energy $\sqrt{s}$ and the PYTHIA tunes considered.   
\end{abstract}

\maketitle

\section{Introduction}
\label{sec:introduction}

Charge balance functions (BFs) were introduced at the beginning of the RHIC era as a tool to investigate the evolution of particle production in heavy-ion collisions~\cite{Bass:2000az,Pratt:2002BFLH,Jeon:2002BFCF} and the presence of delayed hadronization as an indication of the formation of long lived isentropic expanding quark--gluon plasma (QGP) in such high-energy nucleus--nucleus (A--A) collisions. It later emerged that BFs
are rather sensitive to the radial expansion dynamics of the matter formed in A--A collisions~\cite{Voloshin:2006TRE,Pruneau:2007ua,Bozek:2005BFTF,Pratt:2019pnd} and can, in principle, be used to probe hadron production mechanisms such as the formation of clusters~\cite{Pratt:2018ebf}. Later still, it was found that the azimuthal dependence of hadron BFs, particularly heavier hadrons, is sensitive to  the  diffusivity of light quarks~\cite{S.GavinAPHA:2006Diffusion,Jeon:2002BFCF,Pratt:2019pnd}. Furthermore, it has also been argued that general balance functions, i.e., BFs involving distinct species of hadrons, are sensitive to QGP susceptibilities near the phase transition~\cite{Pratt:2015jsa}. Nominally, susceptibilities determine the magnitude of net charge, net strangeness, and net baryon number fluctuations. It is straightforward to show, however, that the magnitude of fluctuations, expressed in terms of net-charge cumulants are strictly related to balance functions and higher order correlation functions~\cite{Pruneau:2019BNC}. Measurements of BFs and other correlation functions offer the distinctive advantage of enabling detailed experimental/instrumental corrections and also allow more comprehensive studies of the impact of final measurement acceptances. It is thus of interest to carry out studies of the feasibility and precision that can be achieved in the pursuit of charge balance functions, baryon balance functions, and strange balance functions. Such exploratory studies must evidently rely on existing models of particle production in elementary collisions (e.g., proton--proton) and heavy-ion collisions. Unfortunately, these studies pose considerable theoretical challenges on the models: to be suitable and workable in the study of charge, baryon, and strangeness balance functions, the models must conserve all these quantum numbers in addition to satisfying basic laws of energy-momentum conservation. Models that satisfy these requirements are few. In this work, we utilize the MONASH tune~\cite{Skands:2014pea} of PYTHIA~\cite{Bierlich:2022pfr}, and two alternative modes known as ``Shoving" and ``Ropes"~\cite{Bierlich:2014xba} to investigate the behavior of light hadron balance functions and baryon balance functions. We are specifically interested in figuring out how the amplitude, and fractional integral of balance functions evolve with beam energy and the specific mechanisms involved in the conversion of partonic degrees of freedom into on-shell hadrons. 

This paper is organized as follows. In Sec.~\ref{sec:Definition} we introduce the unified balance functions~\cite{Pruneau:2022brh} and discuss expectations relative to their integrals. Section~\ref{sec:PYTHIA} presents a brief description of the PYTHIA model and the differences between the MONASH, Ropes, and Shoving modes. Section~\ref{sec:Light Hadron BF} presents studies of light charged hadron balance functions while Sec.~\ref{sec:Baryon BF} discusses studies of baryon balance functions. A summary of this work is presented in Sec.~\ref{sec:summary}. 

\section{Balance Function Definition}
\label{sec:Definition}

In this work, we adopt the unified balance function definition set forth in recent works~\cite{Pruneau:2022brh,Pruneau:2023zhl}. Three bound balance functions, i.e., balance function determined within a limited acceptance $\Omega$, are nominally defined. The first  represents the (conditional) density of particles of type $\alpha$ being detected at rapidity $y_1$ when an anti-particle of type $\bar\beta$ is found at rapidity $y_2$. The second represents the complement of the first, i.e., the density of detecting an anti-particle $\bar\alpha$ at $y_1$ when a particle of type $\beta$ is found at $y_2$. The third is simply the average of the first two. They are defined as
\begin{align}
\label{eq:B2alphaBarBetay1y2}
   B^{\alpha\bar\beta}(y_1,y_2|\Omega) &= 
   \frac{1}{\langle N_1^{\bar\beta}\rangle}\left[ C_2^{\alpha\bar\beta}(y_1,y_2) - C_2^{\bar\alpha\bar\beta}(y_1,y_2) \right], \\
\label{eq:B2BarAlphaBetay1y2}
   B^{\bar\alpha\beta}(y_1,y_2|\Omega) &=    \frac{1}{\langle N_1^{\beta}\rangle}\left[ C_2^{\bar\alpha\beta}(y_1,y_2) - C_2^{\alpha\beta}(y_1,y_2) \right], \\
\label{eq:Bs}
   {B}^{\alpha\beta,s}(y_1,y_2|\Omega) &=\frac{1}{2}\left[ B^{\alpha\bar\beta}(y_1,y_2|\Omega)
   +B^{\bar\alpha\beta}(y_1,y_2|\Omega)\right],
\end{align}
where $C_2^{\alpha\beta}(y_1,y_2)$ are two-particle differential cumulants computed according to
\begin{equation}
\label{eq:C2}
    C_2^{\alpha\beta}(y_1,y_2) = \rho_2^{\alpha\beta}(y_1,y_2) - \rho_1^{\alpha}(y_1)\rho_1^{\beta}(y_2),
\end{equation}
in which $\rho_1^{\alpha}(y_1)$ and $\rho_1^{\beta}(y_2)$ are single particle densities of species $\alpha$ and $\beta$, respectively, and $\rho_2^{\alpha\beta}(y_1,y_2)$ is the (joint) pair density of these two species. All densities are determined within the fiducial acceptance $\Omega$. The quantities $\langle N_1^{\alpha}\rangle$ and
$\langle N_1^{\beta}\rangle$ are the event ensemble averages of the number of particles of types $\alpha$ and $\beta$ obtained on an event-by-event basis within $\Omega$. The functions  $B^{\alpha\bar\beta}(y_1,y_2|\Omega)$ and $B^{\bar\alpha\beta}(y_1,y_2|\Omega)$ nominally yield different dependencies on the rapidities. They are of interest to examine differences in charge balancing of particles of type $\beta$ and their anti-particle $\bar\beta$. In this work, however, we focus on the average balancing of charges (and baryon number) and consider mixed balance function computed with Eq.~(\ref{eq:Bs}) exclusively. Additionally, given that our goal is to study the sensitivity of mixed species of balance functions, and their integrals, the functions are computed based on the pair rapidity difference $\Delta y=y_1-y_2$ according to
\begin{align}
    B^{\alpha\beta,s}(\Delta y) &\equiv   \int_{\Omega} {\rm d}\bar y \hspace{0.05in} B^{\alpha\beta,s}(\Delta y,\bar y). 
\end{align}
This enables a straightforward study of integrals of $B^{\alpha\beta,s}(\Delta y)$, denoted herewith $I^{\alpha\beta,s}(\Delta y)$
\begin{align}
\label{eq:CumulativeIntegralBF}
    I^{\alpha\beta,s}(\Delta y) &\equiv   \int_{-\Delta y}^{\Delta y} {\rm d}\Delta y' \hspace{0.05in} B^{\alpha\beta,s}(\Delta y').
\end{align}
Given a particle of type $\bar{\beta}$, charge (or baryon number) balancing requires that the sum of all balance functions $B^{\alpha\beta,s}(\Delta y)$ spanning particle types $\alpha$ of opposite charge (or baryon number) must integrate to unity under a full acceptance condition. In the following, for simplicity, these are denoted as $\sum I^s$ and values of these sums (of integrals of $B^{\alpha\beta,s}(\Delta y)$) are studied as a function of the breadth $\Delta y$ of the acceptance. 

It is important to note that balance functions and cumulative integrals presented in this work are computed based on an ideal and perfect acceptance (i.e., $p_{\rm T}>0$ and $4\pi$ coverage) to focus the discussion on limiting integrals of balance functions and their general evolution with the width of the longitudinal acceptance. In actual experiments, the acceptance is usually limited longitudinally to a range $-y_0 \le y < y_0$ and one cannot sample the full phase space of particle production. This leads to ``diamond shaped" particle pair acceptances when 
expressed in terms of the pair rapidity difference $\Delta y=y_1 -y_2$ and the rapidity average $\bar{y} = (y_1+y_2)/2$. The narrowing of the acceptance for increasing values of $\Delta y$ can, however, be partially compensated for by diving the strength of correlation functions at a given value of $\Delta y$ by the width of the $\bar y$ acceptance $1-\Delta y/2y_0$. Such compensation is not considered in this work.

It is also of interest to note that balance functions can be formulated for all conserved quantum numbers, i.e., including strangeness, charm, bottomness, etc, in addition to the electric charge and baryon number considered in this work. In this context, one must remark that if elementary particles (and nuclei) with multiple units of charge, strangeness, etc, were considered, the balance functions would need to be computed based on the number of units of charge, strangeness, etc carried by the particles: the yields would need to be multiplied by the number of units of charge, strange quarks, etc, carried by each particle.

\section{Simulation Model -- PYTHIA8}
\label{sec:PYTHIA}

The charge BFs of particles produced in 
high-energy proton--proton (pp) collisions are investigated theoretically based on simulations carried out with the PYTHIA8  Monte Carlo event generator operated with the MONASH 2013 tune~\cite{Skands:2014pea} with color reconnection, as well as with the so called Shoving and Ropes tunes~\cite{Bierlich:2014xba, Bierlich:2018lbp}.
PYTHIA8 is based on a QCD description of quark and gluon interactions at leading order (LO) and uses the Lund string fragmentation model for high-$p_{\rm T}$ parton hadronization while the production of soft particles (i.e., the underlying event) is handled through fragmentation of mini-jets from initial and final state radiation, as well as multiple parton interactions~\cite{Sjostrand:2007gs}. The Shoving and Ropes tunes are respectively designed to illicit flow-like behavior and increase the production of higher particle mass, specifically strange mesons and baryons.  

Studies reported in this work are carried with PYTHIA running in minimum-bias mode, with soft QCD processes and color reconnection turned ON. The Shoving and Ropes modes use  parameter values prescribed by current developers of PYTHIA~\cite{Bierlich:2018lbp} (https://gitlab.com/Pythia8).

Events were generated and analysed on the Wayne State University and Institute of Space Science (Romania) computing grids in groups of 10 or more jobs, each with 10 or more sub-jobs, and 300,000 events per sub-job. This enabled efficient and rapid use of the grids as well as reliable statistical uncertainty calculations based on the sub-sample technique.

\section{Light Hadron Charge Balance Functions}
\label{sec:Light Hadron BF} 

The overall shape of charge balance functions is largely driven by the processes that lead to the production and transport of particles. Indeed, in addition to string fragmentation, particlization, and decays, one should also note the role of radial flow~\cite{Pratt:2018ebf,Voloshin:2002ku} and the diffusion of light quarks~\cite{Pratt:2019pnd,Pratt:2021xvg}, amongst others. One thus expects that within a given acceptance in rapidity, these different processes should influence the integral as well as the shape of the BFs. Measurements of the relative contributions and integrals to charge balancing should thus provide new discriminating power over the many models of particle production~\cite{PhysRevC.99.044916,ALICE:2021hjb}. It is thus worth pursuing measurements of mixed and same species balance functions in several distinct transverse momentum and rapidity ranges. 

Experimentally, measurements of identified hadrons $\pi^{\pm}$, ${\rm K}^{\pm}$, and $\rm p/\bar p$ are by far the most straightforward. Technological advances achieved with modern experiments, however, also enable precise measurements of strange baryons, D-mesons, and B-mesons. It is thus interesting to consider mixed  balance functions of these particles also. Predictions of balance functions of mixed low mass charge hadrons and baryons, based on PYTHIA, are presented in this and the next section whereas those of D-mesons and B-mesons are left for future studies. 

At the Large Hadron Collider (LHC), measurements of single particle spectra of neutral and charged pions ($\pi^{\pm,0}$), kaons (${\rm K}^{\pm,0}$), as well as protons/anti-protons ($\rm p, \bar p$) are quantitatively well reproduced by models such as PYTHIA~\cite{Skands:2014pea}, EPOS4~\cite{Werner:2023jps}, and hydrodynamical based models~\cite{Du:2023gnv,Nijs:2020roc,Putschke:2019yrg,Bleicher:1999xi}. In these models, sets of high-mass hadrons are usually explicitly included with in-vacuum decay modes to simulate their decay and the production of low mass hadrons (e.g., $\pi^{\pm,0}$, ${\rm K}^{\pm,0}$, $\rm p$, $\rm \bar p$, and so on). While little or no ambiguity exists in calculations based on PYTHIA and transport models~\cite{Bleicher:1999xi},  uncertainties do arise, in general, in the context of hydrodynamics based models because of the intricacies associated with the particlization of the freeze out surface of the energy-momentum tensors and from contributions of jet fragmentation. It is then legitimate to ask what fraction of low mass hadrons originate from decays of high mass states relative to ``direct production" by fragmentation of strings or jets or the hadronization of the QGP. It is then of interest to consider the joint (i.e., correlated) production of mixed pairs of particle species to establish the relative yield of correlated particle production. This can be done unambiguously based on mixed species balance functions. Indeed, as illustrated below, the strength and shape of mixed species balance function are directly sensitive to the production processes that lead to joint production of charge (baryon number) balancing hadrons. Additionally, balancing functions present the singular advantage, relative to more generic two-particle correlation functions, of featuring a simple sum rule: the sum of integrals $I^{\alpha\beta,s}$ must converge to unity for measurements within a full acceptance (i.e., $4\pi$ acceptance and no losses at low or high $p_{\rm T}$). It is thus meaningful to experimentally study what values $I^{\alpha\beta,s}$ are produced by pp, p--A, and A--A collision systems and to require these values be well reproduced by simulation models. 

In the context of PYTHIA, the rapidity and transverse momentum of particles are determined in part by the string fragmentation and color reconnection processes, as well as by the decay of short-lived hadrons. Such decays are constrained by the mass of the parent and daughter particles and should thus yield balance functions that are sensitive to the kinematics of the parent particles. Given that decays usually yield particles at the low end of the transverse momentum spectrum, one then expects balance functions to feature great sensitivity to the spectrum of the parent hadrons as well as the relative abundances of both light and heavy hadrons. Evidently, correlations may also arise from the underlying particle production mechanisms, whether associated with the fragmentation of strings or jets or the particlization of a radially flowing freeze out surface. Measurements of mixed balance functions and their evolution with transverse momentum and rapidity shall then provide new additional information that will improve the understanding of particle production and the matter they emerge from. We first illustrate this point with calculation of mixed light hadron balance functions.  

\begin{figure}[ht]
 	\includegraphics[width=0.32\linewidth,trim={12mm 1mm 19mm 3mm},clip]
    {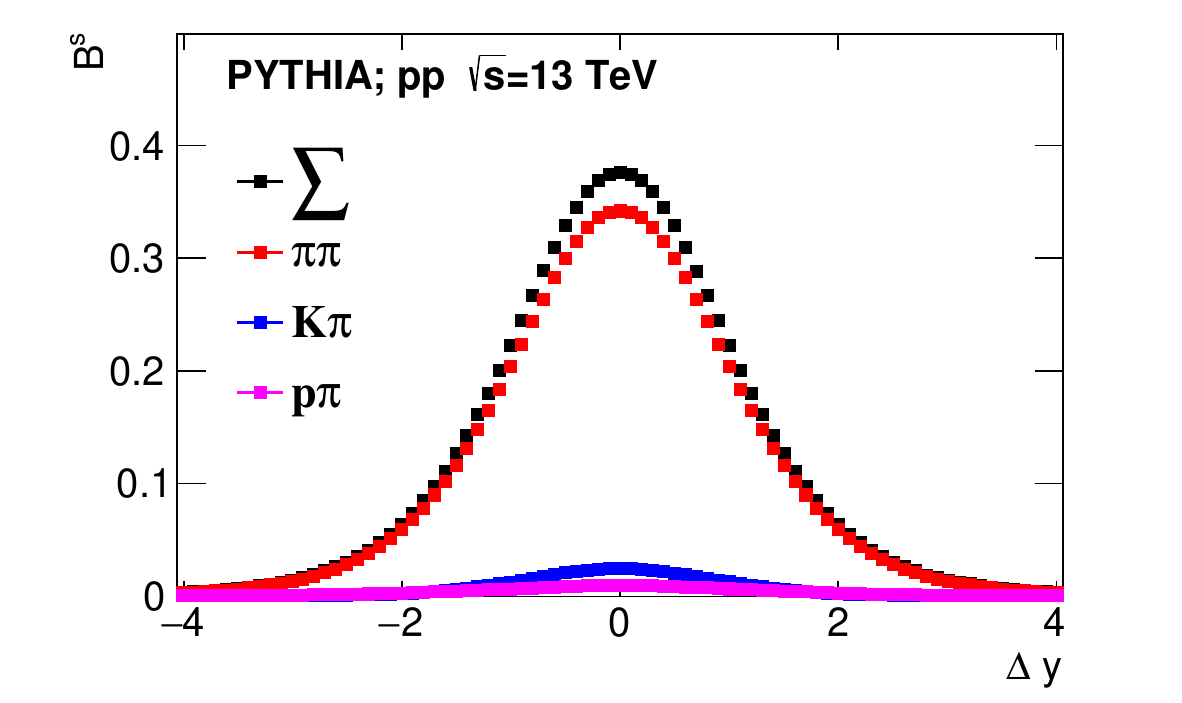}
 	\includegraphics[width=0.32\linewidth,trim={12mm 1mm 19mm 3mm},clip]
    {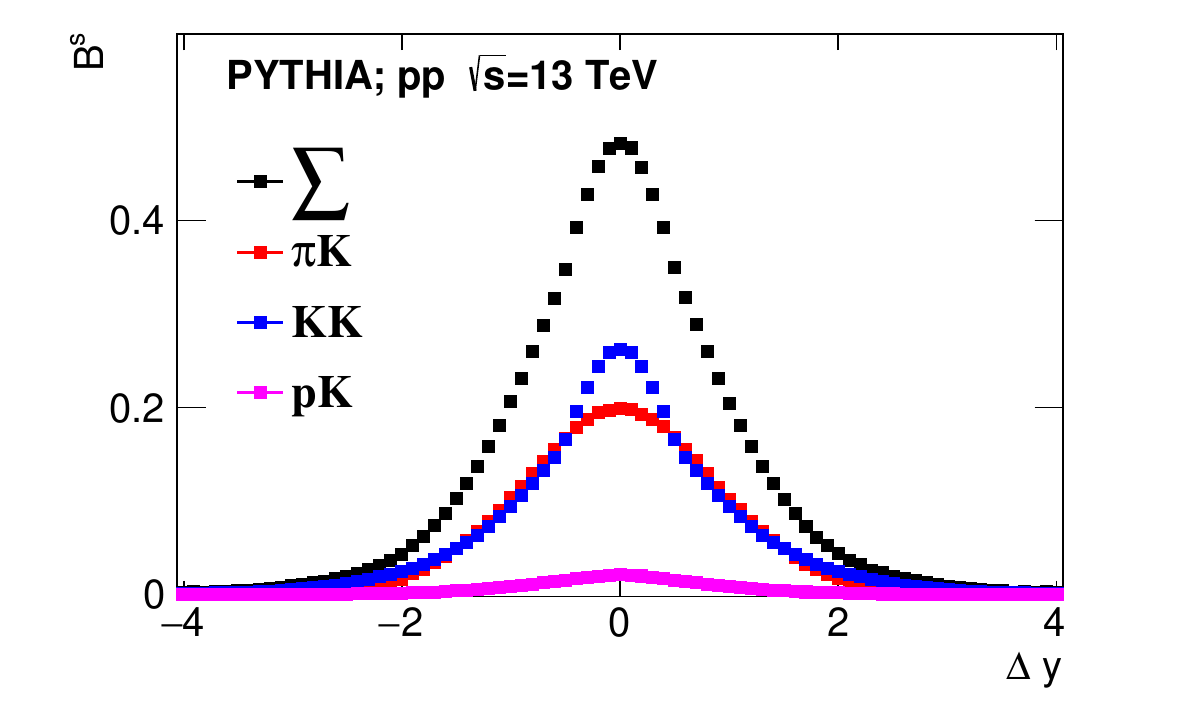}
 	\includegraphics[width=0.32\linewidth,trim={12mm 1mm 19mm 3mm},clip]
    {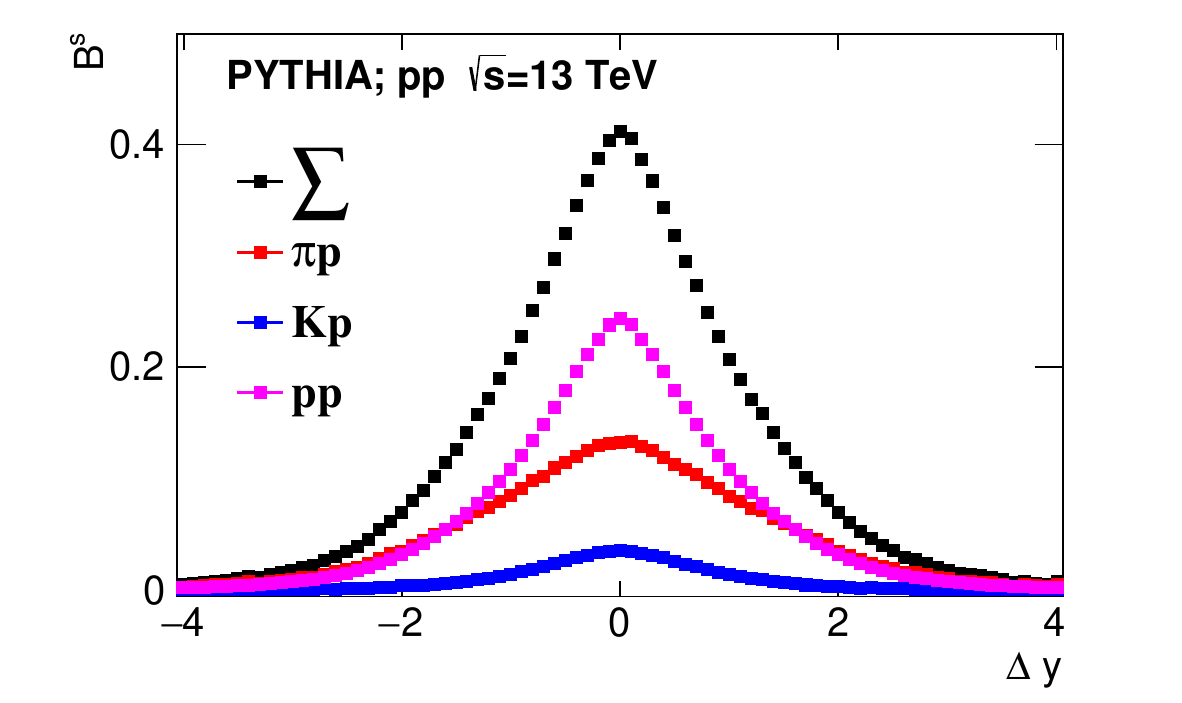}
	%	Is
 	\includegraphics[width=0.32\linewidth,trim={12mm 1mm 19mm 3mm},clip]
    {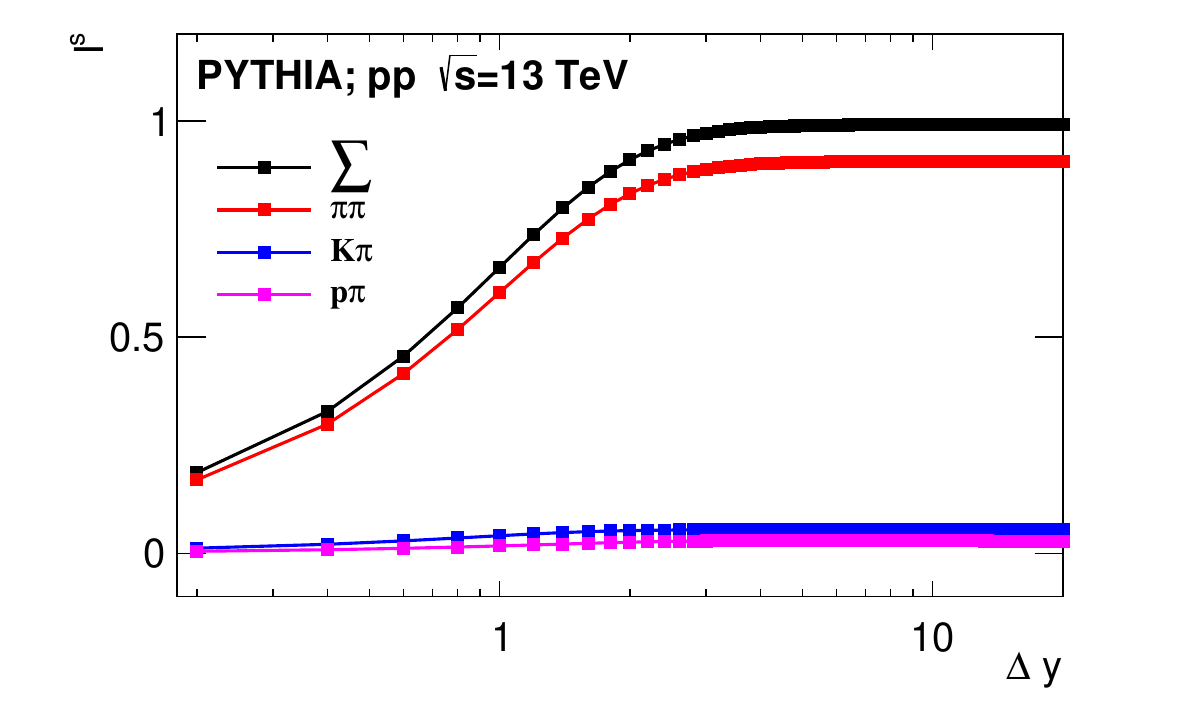}
 	\includegraphics[width=0.32\linewidth,trim={12mm 1mm 19mm 3mm},clip]
    {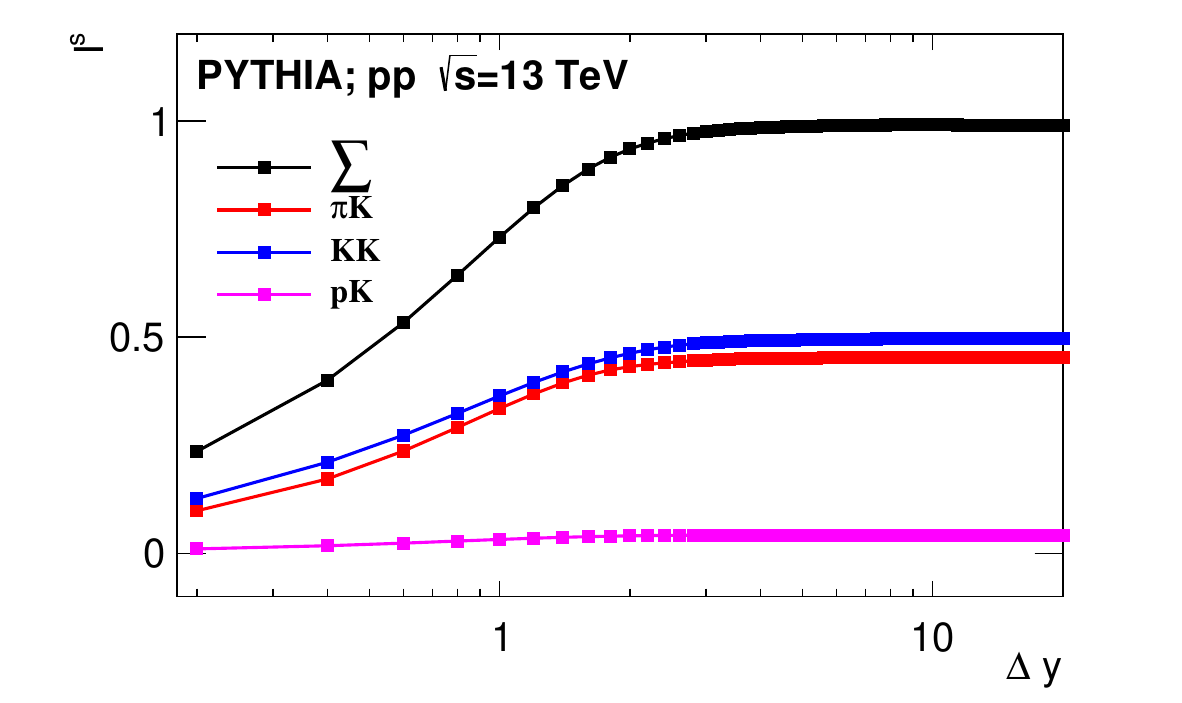}
 	\includegraphics[width=0.32\linewidth,trim={12mm 1mm 19mm 3mm},clip]
    {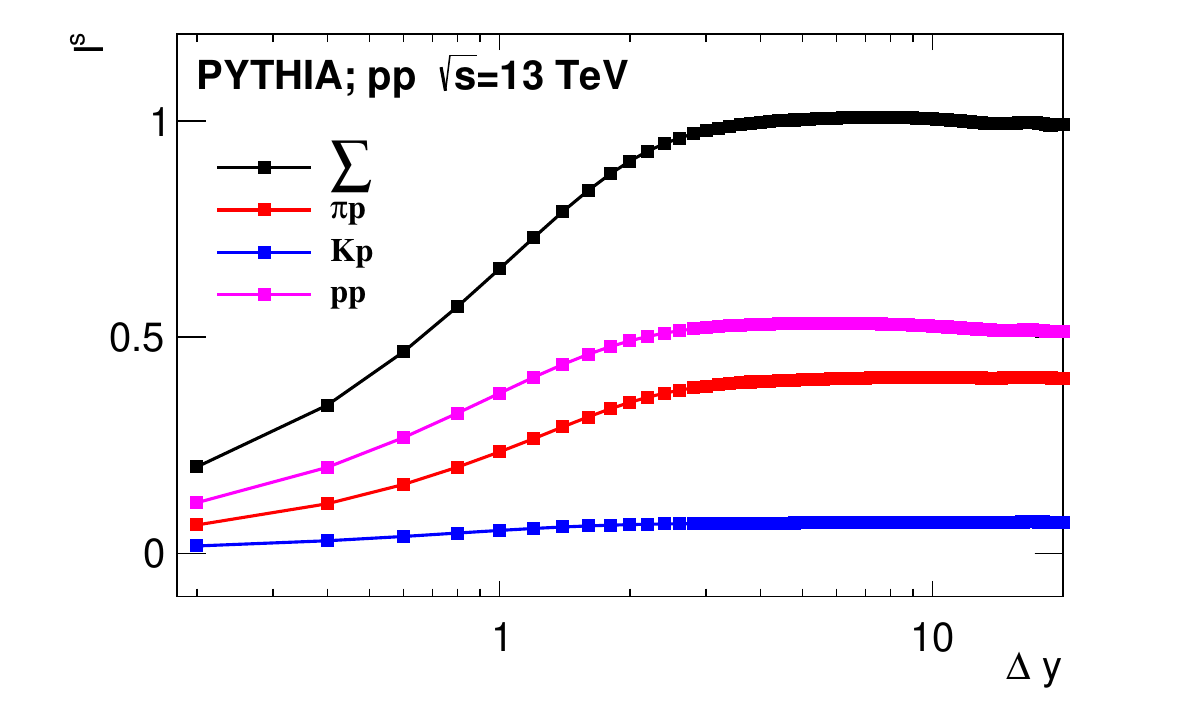}
 \caption{(top) Mixed balance functions $B^{\alpha\beta,s}$ 
 of charged light hadrons $\alpha,\beta=\pi^{\pm}$, ${\rm K}^{\pm}$, $\rm p(\bar{p})$ and (bottom) their respective cumulative integrals $I^{\alpha\beta,s}$, computed for pp collisions at $\sqrt{s}=13$ TeV with PYTHIA8 Monash tune with color reconnection. The second label of a pair $\alpha\beta$ identifies the ``reference" particle, whereas the first label (e.g., $\alpha$) indicates the balancing partner. The symbol $\sum$ identifies the sum of balance functions (top) and their integrals (bottom). See text for details.}
\label{fig:piKpBF}
\end{figure}

The top panels of Fig.~\ref{fig:piKpBF} present mixed species balance functions of pairs $\pi^{\mp}\pi^{\pm}$, ${\rm K}^{\mp}\pi^{\pm}$,
and ${\rm \bar p}\pi^+$ (${\rm p}\pi^-$)
computed according to Eq.~(\ref{eq:Bs}) from  pp collisions at $\sqrt{s}=13$ TeV simulated with PYTHIA8 (MONASH tune) with color reconnection. The bottom panels display cumulative integrals $I^{\alpha\beta,s}(\Delta y)$, computed according to Eq.~(\ref{eq:CumulativeIntegralBF}). By convention, in this and following figures, the second label of a pair $\alpha\beta$ identifies the ``reference" particle, whereas the first label (e.g., $\alpha$) indicates its charge balancing partner. Weak decays of light hadrons produced by PYTHIA8 (e.g., ${K_S^0}$, $\Lambda^0$, etc) are turned off in order to focus the calculation on balance functions of primary particles. 

One finds that the amplitude of $\pi^{\mp}\pi^{\pm}$ BF (top/left panel) significantly dominates that of the ${\rm K}^{\mp}\pi^{\pm}$,
and ${\rm \bar p}\pi^+$ (${\rm p}\pi^-$) BFs. This is also clear from the bottom/left panel which displays the cumulative integral $I^{\alpha\pi^{\pm},s}(\Delta y)$. One observes in particular that the three BF integrals $I^{\alpha\pi^{\pm},s}(\Delta y)$ (i.e., triggered by $\pi^{\mp}\pi^{\pm}$) rapidly rise in the range $0<\Delta y<1.5$ and eventually saturate for $\Delta y>3$. Clearly, the integral of the $\pi^{\mp}\pi^{\pm}$ BF far out weights those of ${\rm K}^{\mp}\pi^{\pm}$,
and ${\rm \bar p}\pi^+$ (${\rm p}\pi^-$) BFs. This implies that 
the charge of a $\pi^+$ ($\pi^-$) is most likely balanced by the production of $\pi^-$ ($\pi^+$), while balancing by production of ${\rm K}^-$ (${\rm K}^+$) or ${\rm \bar p}$ (${\rm p}$) is possible but far less probable. Additionally note, based on the bottom/left panel of Fig.~\ref{fig:piKpBF}, that the integrals 
$I^{\pi^{\mp}\pi^{\pm},s}(\Delta y)$, 
$I^{{\rm K}^{\mp}\pi^{\pm},s}(\Delta y)$, and $I^{{\rm p(\bar p)}\pi^{\pm},s}(\Delta y)$ add to unity in the large acceptance limit, $\Delta y\rightarrow 20$. This indicates, in the context of the PYTHIA model, that the production of a charged pion is (nearly) always charge balanced by the production of a charged pion, kaon, or proton of the opposite charge. While the charge balancing could in principle be accomplished by the presence of electrons ($e^{\pm}$) resulting from decays of heavy quarks, the probability  of such processes is rather small and thus does not appreciably contribute to  the sum $\sum_{\alpha}I^{\alpha\pi^{\pm},s}(\Delta y)$. 
It should be noted once again that cumulative integrals $I^{\alpha\pi^{\pm},s}(\Delta y)$ were obtained from BFs computed in full acceptance $|y|= 10$ and as such may not accurately reflect values that might be obtained with smaller experimental acceptances. 

Based on central panels of Fig.~\ref{fig:piKpBF}, one observes that the charge balancing of ${\rm K}^{+}$ proceeds somewhat differently than the balancing of $\pi^{+}$. Note indeed that the production of a ${\rm K}^{+}$ is almost as often accompanied by a $\pi^{-}$ as ${\rm K}^{-}$, but is much less likely to be balanced by a $\rm \bar p$. Charge balancing by emission of ${\rm K}^{-}$ also satisfies the strangeness balance of the ${\rm K}^{+}$ but charge balancing by a $\pi^{-}$ requires a third (neutral) particle be emitted (and causally connected) to balance the strangeness of the ${\rm K}^{+}$. Similarly, charge balancing by production of a $\rm \bar p$, a non-strange baryon, requires at least one additional particle is produced, e.g., some strange neutral baryon in order to balance both the strangeness of the ${\rm K}^{+}$ and baryon number of the $\rm \bar p$. Additionally note that the charge balancing of the ${\rm K}^{+}$ by 
$\pi^{-}$ as ${\rm K}^{-}$ and $\rm \bar p$ effectively add up to unity: processes yielding negative heavy mesons decaying with the emission of an electron have small cross section and contribute negligibly to the charge balancing of the ${\rm K}^{+}$.

The charge balancing of proton (${\rm p}$), illustrated in right panels of Fig.~\ref{fig:piKpBF}, is also of particular interest. One finds that the charge of a ${\rm p}$ is most often balanced by the charge of an anti-proton (likely by a pair creation process that also balances the baryon number of the proton) but balancing by the production of a $\pi^{-}$ is also very probable while balancing by emission of a ${\rm K}^{-}$ is less probable but nonetheless possible. The latter must involve the emission of a least one more causally connected particle, e.g., a $\Lambda^0$, to also balance strangeness and baryon quantum numbers. As such, this implies that balance functions of pairs such as $\pi\rm K$ or $\rm pK$ probe one facet of more complex processes involving conservation of charge, strangeness, and baryon number. Mixed species balance functions may thus probe complex processes involving the production of three or more particles constrained by two or more quantum number conservation laws. 

%{\color{blue}
%It is also of interest to remark that the $\rm p\bar p$ balance function is suppressed (e.g., negative) at very large rapidity difference. This likely corresponds to processes in which the baryon number and charge of an incoming proton is transferred to other produced particles thereby leading to a deficit of charge/baryon number balancing at near-beam rapidity. While such suppression would be hard to observe in practical experimental conditions, the deficit at large $\Delta y$ is found to be partly compensated in the rapidity $4 < |\Delta y| < 10$ where the BF is slightly above zero. Such modest positive value might be observable with very wide acceptance spectrometer and would thereby constitute a mechanism to probe the unobservable suppression at very large $\Delta y$.}

\begin{figure}[tp]
 	\includegraphics[width=0.32\linewidth,trim={4mm 1mm 15mm 3mm},clip]
    {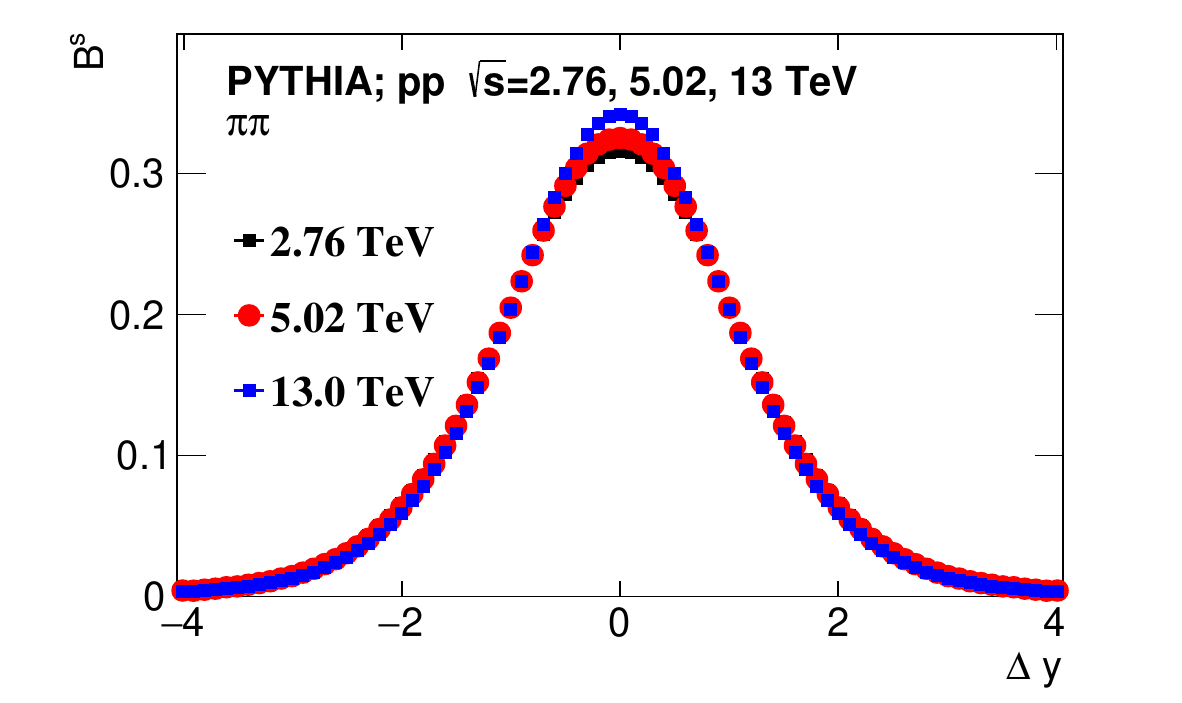}
 	\includegraphics[width=0.32\linewidth,trim={4mm 1mm 15mm 3mm},clip]
    {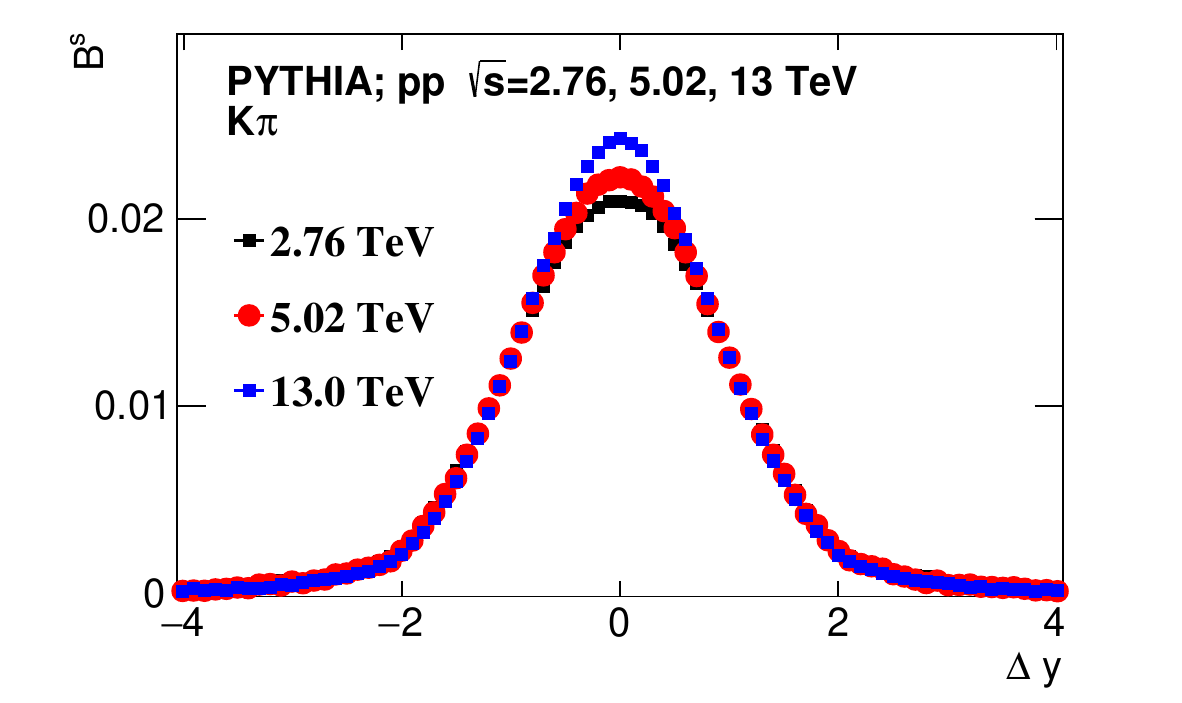}
 	\includegraphics[width=0.32\linewidth,trim={4mm 1mm 15mm 3mm},clip]
    {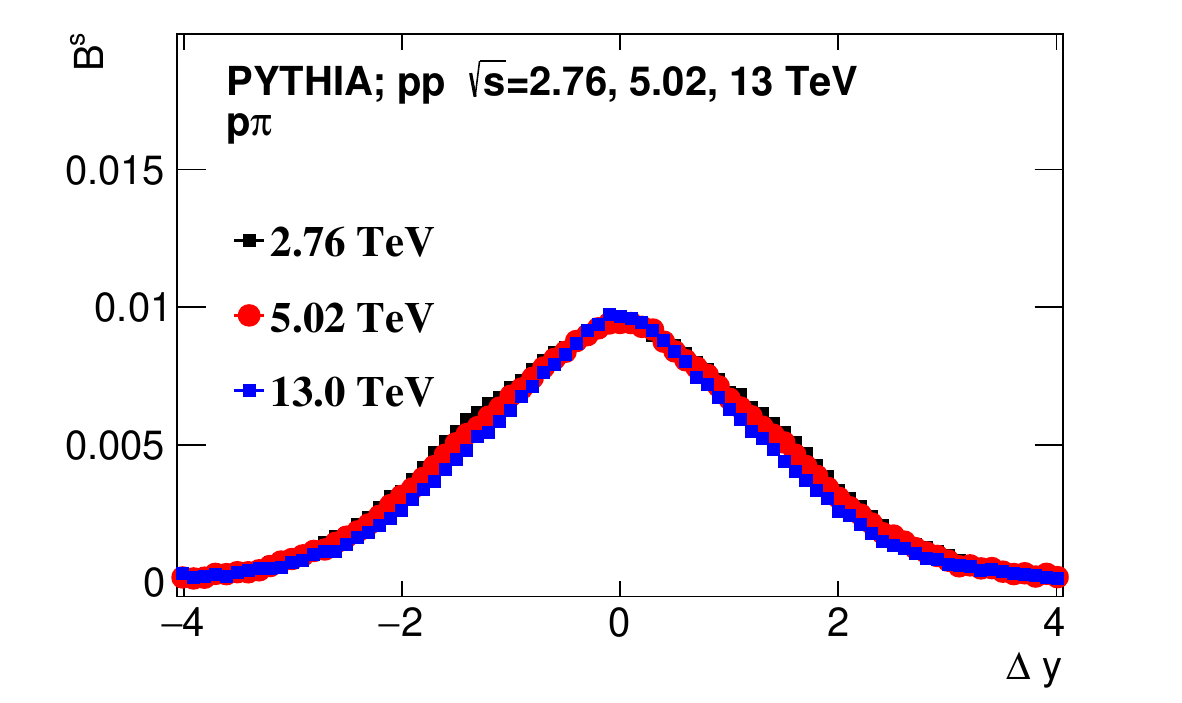}
 	\includegraphics[width=0.32\linewidth,trim={4mm 1mm 15mm 3mm},clip]
    {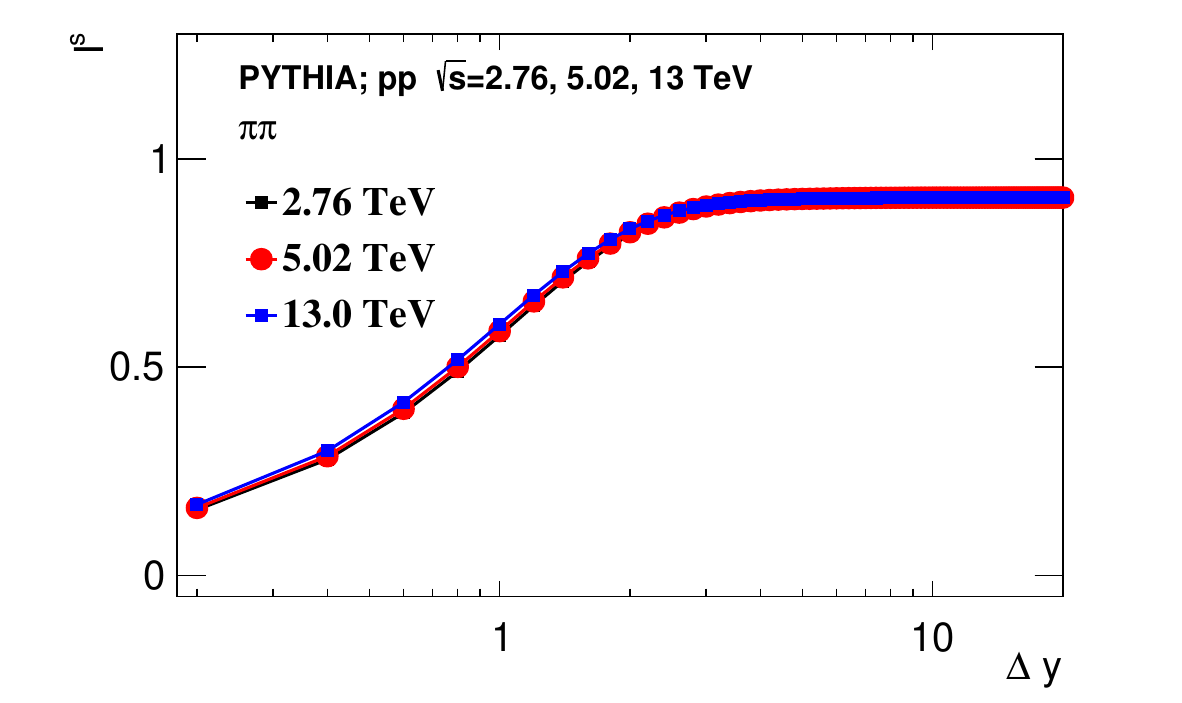}
 	\includegraphics[width=0.32\linewidth,trim={4mm 1mm 15mm 3mm},clip]
    {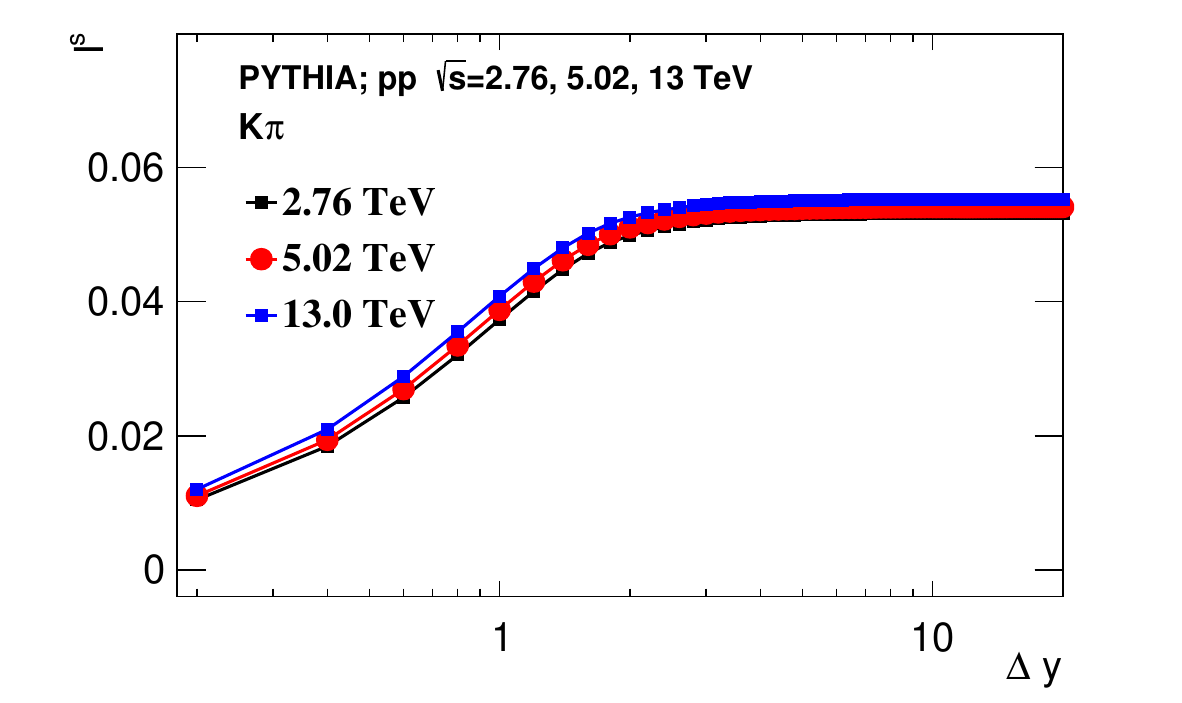}
 	\includegraphics[width=0.32\linewidth,trim={4mm 1mm 15mm 3mm},clip]
    {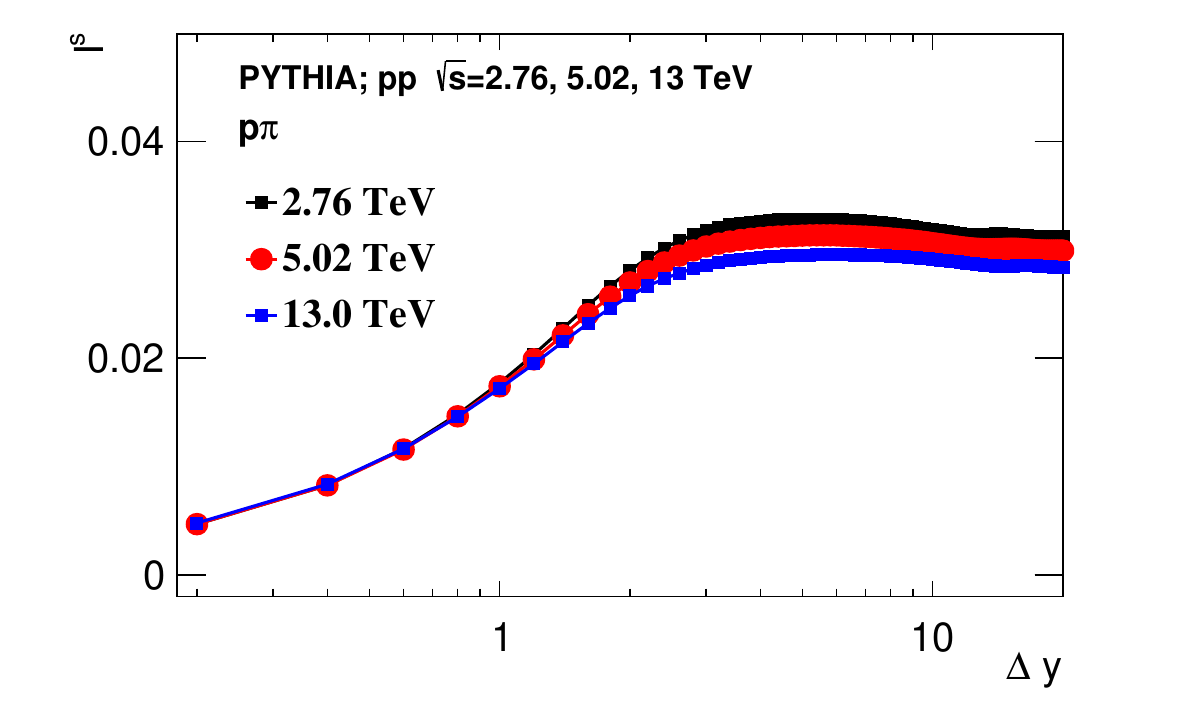}
\caption{(top) Selected mixed pair balance functions $B^{\alpha\beta,s}$ and (bottom) their cumulative integrals $I^{\alpha\beta,s}$ computed for pp collisions at $\sqrt{s}=2.76$, $5.02$, and $13$ TeV with PYTHIA8 Monash tune with color reconnection.}
	\label{fig:piKpBvsS}
\end{figure}

Balance functions and their integrals, shown in Fig.~\ref{fig:piKpBF}, characterize the charge balancing processes expected when using PYTHIA for pp collisions at $\sqrt{s}=13$ TeV. Given the collision energy ($\sqrt{s}$) determines the relative abundance of hard (jets) and soft particle processes, one should expect that the charge balance of light hadrons might slowly evolve with rising $\sqrt{s}$. 
Such possibility is examined quantitatively in Fig.~\ref{fig:piKpBvsS} which displays selected balance functions (top) and their integrals (bottom) computed with PYTHIA8 Monash tune with color reconnection for $\sqrt{s}=2.76$, $5.02$, and $13$ TeV.  
The BFs exhibit different levels of sensitivity to collision energy. For instance, the shape of the $\pi^{\mp}\pi^{\pm}$ BFs is found to exhibit a small sensitivity to rising $\sqrt{s}$: its amplitude at $\Delta y=0$ slowly rises with $\sqrt{s}$ while its width exhibits a modest reduction. The ${\rm K}^-\pi^+$ BF exhibits a similar increase of its amplitude (and associated decrease of its width) with rising 
$\sqrt{s}$ and one also observes that its integral shows a modest increase with increasing $\sqrt{s}$ whereas the $\pi^-\pi^+$ BF integral does not exhibit any significant increase in the limit $\Delta y\rightarrow 20$. 
The amplitude of the ${\rm p}\pi$ BF, in contrast, does not feature an appreciable dependence on $\sqrt{s}$ but its shape does and, as a by product of this dependence, 
the integral $I^{{\rm p}\pi,s}$ decreases in magnitude at $\Delta y\rightarrow 20$ with rising $\sqrt{s}$ and thus compensates for the observed increase of $I^{{\rm K}\pi,s}$. 
\begin{figure}[!ht]
	%	\centering
\includegraphics[width=0.32\linewidth,trim={4mm 1mm 15mm 3mm},clip]
    {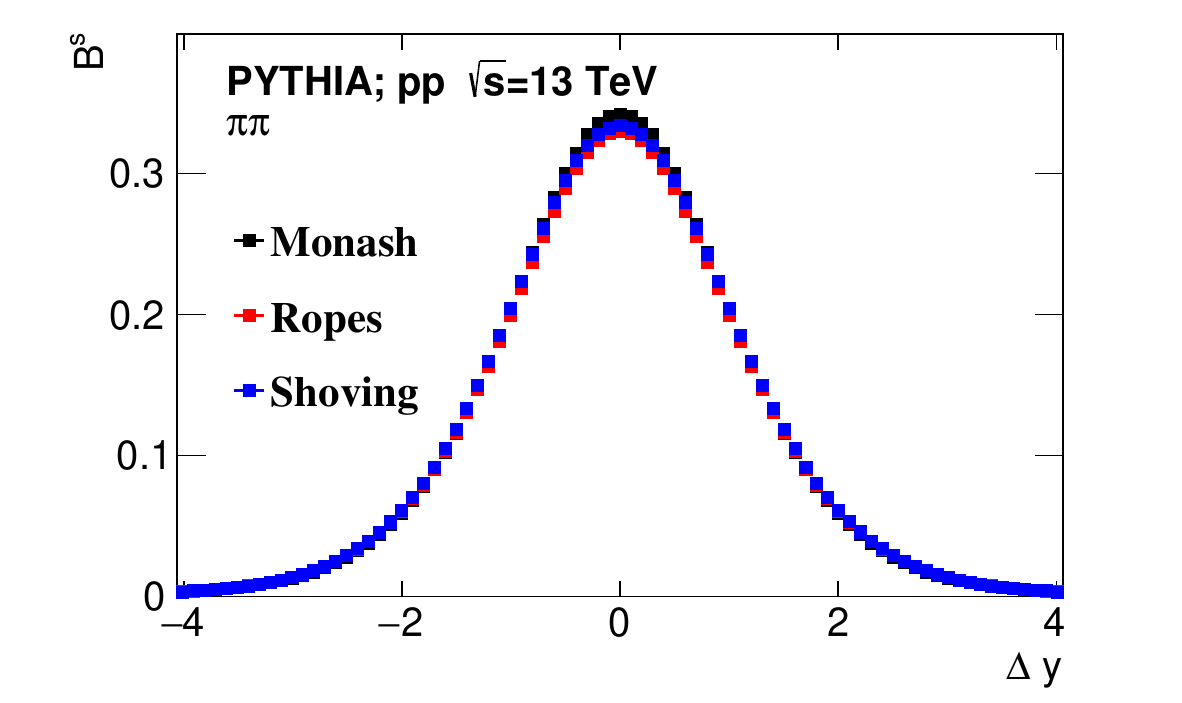}
\includegraphics[width=0.32\linewidth,trim={4mm 1mm 15mm 3mm},clip]
    {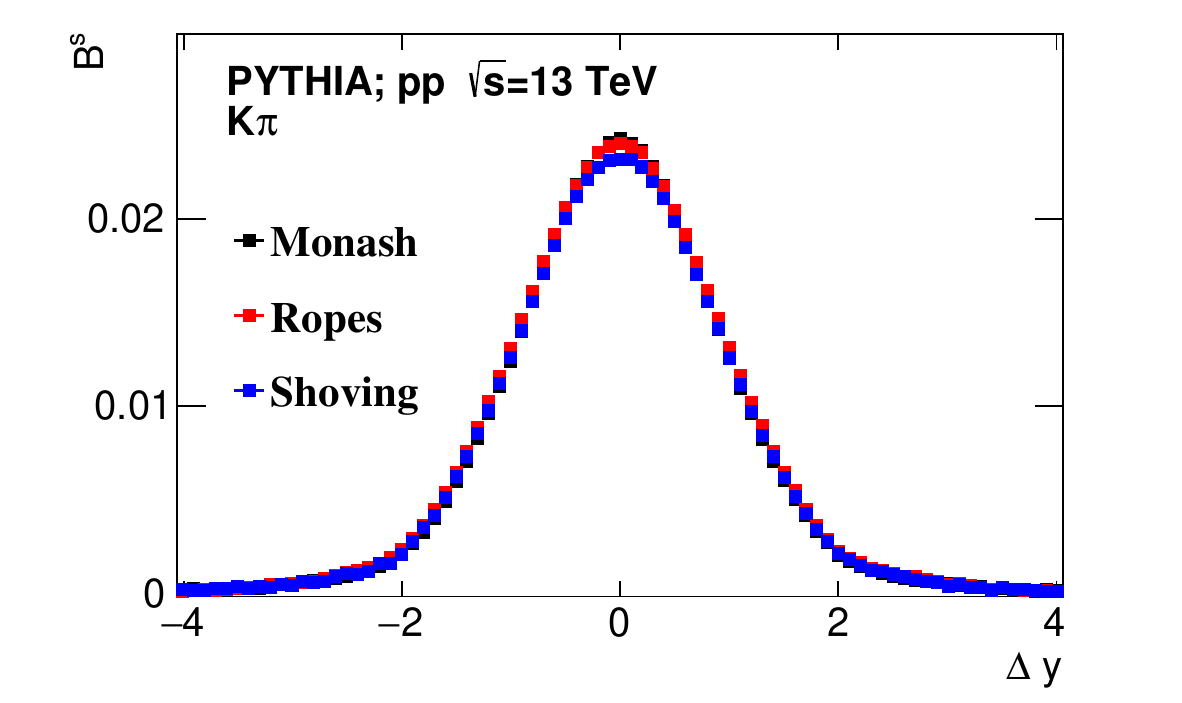}
\includegraphics[width=0.32\linewidth,trim={4mm 1mm 15mm 3mm},clip]
    {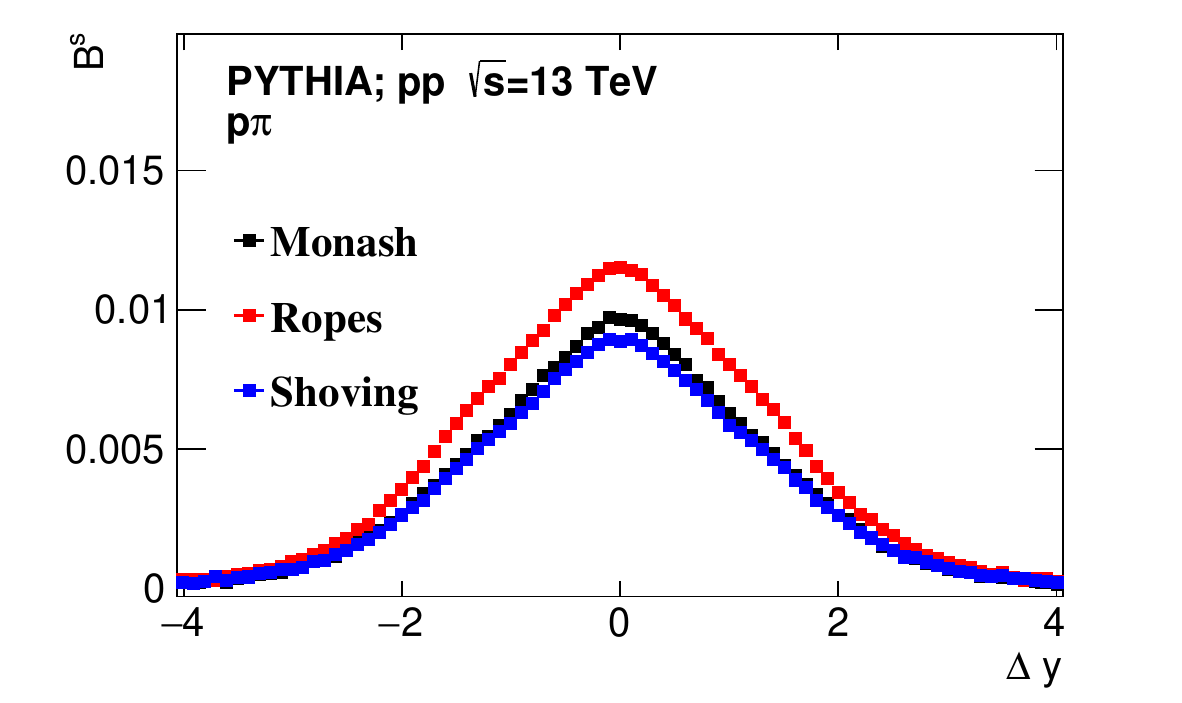}
\includegraphics[width=0.32\linewidth,trim={4mm 1mm 15mm 3mm},clip]
    {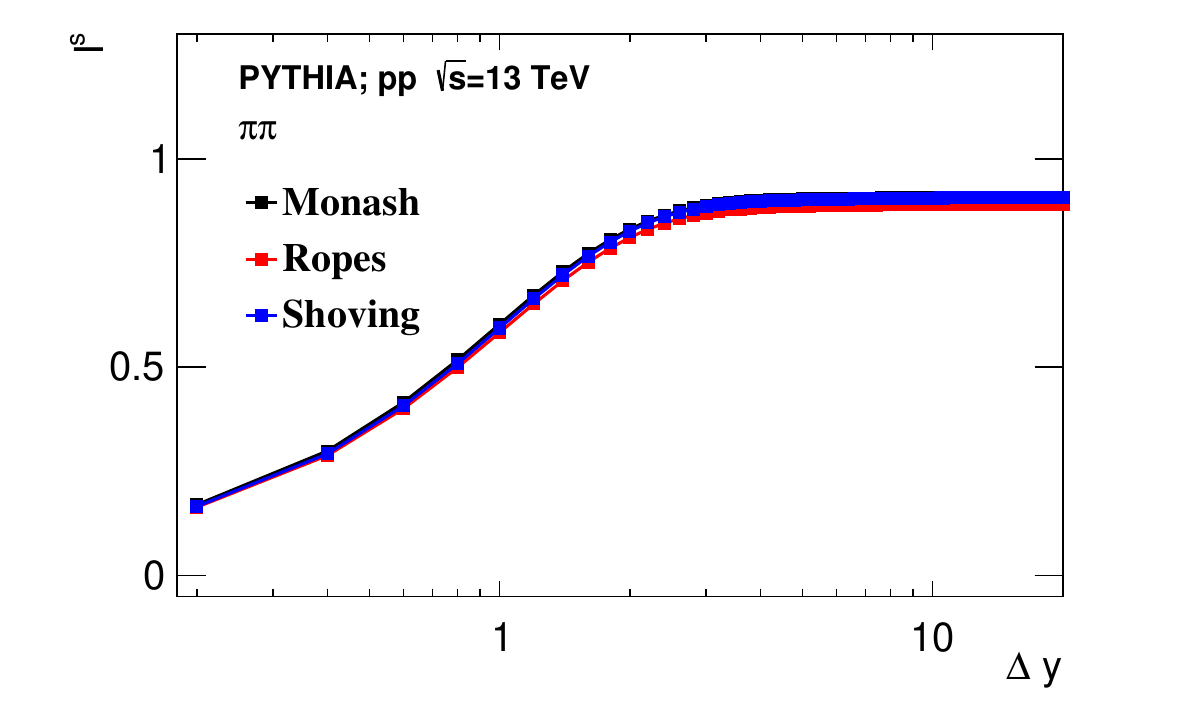}
\includegraphics[width=0.32\linewidth,trim={4mm 1mm 15mm 3mm},clip]
    {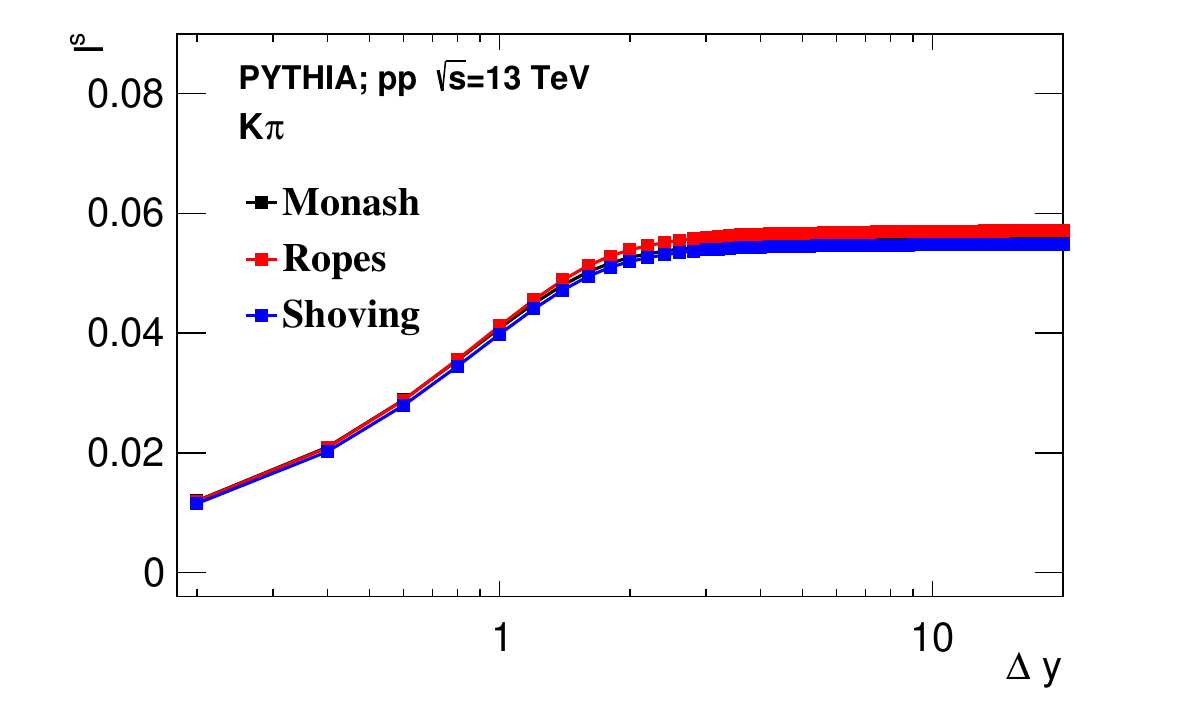}
\includegraphics[width=0.32\linewidth,trim={4mm 1mm 15mm 3mm},clip]
    {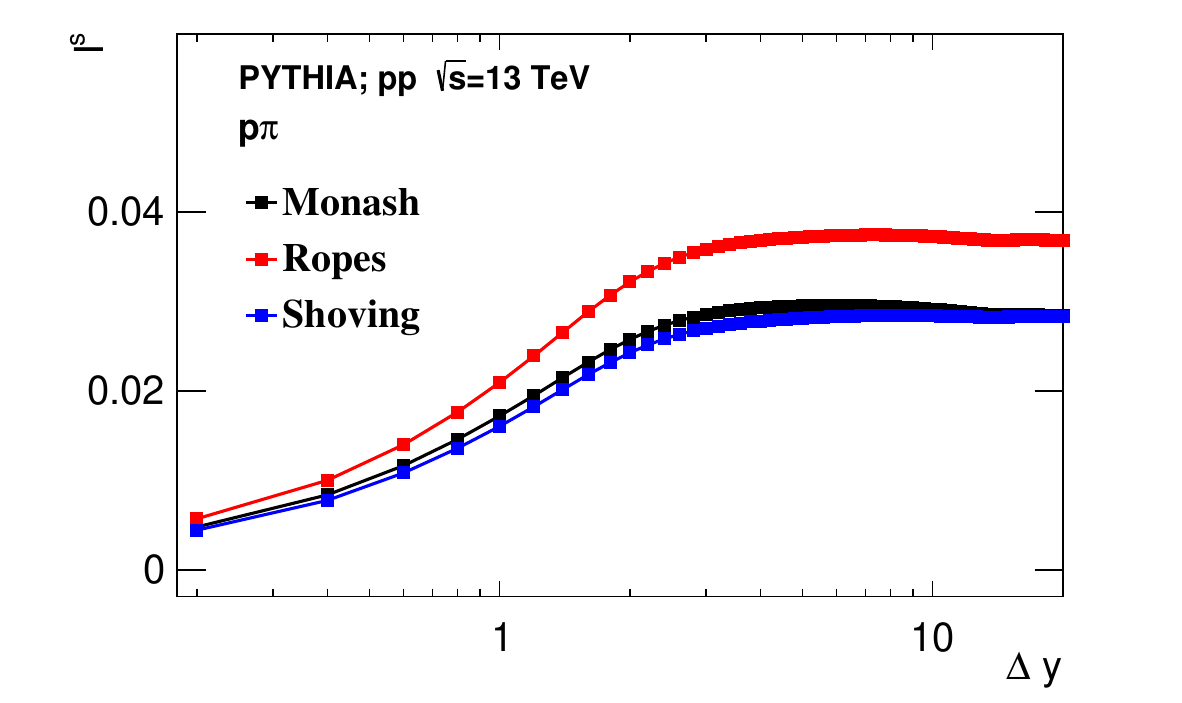}
\caption{(top) Selected mixed pair balance functions $B^{\alpha\beta,s}$ and (bottom) their cumulative integrals $I^{\alpha\beta,s}$ computed for pp collisions at $\sqrt{s}=13$ TeV with the PYTHIA8 Monash tune and Ropes and Shoving modes (with color reconnection). }
	\label{fig:piKpBsVsTunes}
\end{figure}
It is also of interest to examine how qualitative changes in the particle production may impact BF shapes and the saturation value of their integrals. To this end, Fig.~\ref{fig:piKpBsVsTunes} compares selected mixed pair BFs, and their respective integrals, obtained with PYTHIA8 simulations of pp collisions at $\sqrt{s}=13$ TeV performed with the Monash tune and the Ropes and Shoving modes. The shape and amplitude of BFs of pairs $\pi\pi$, ${\rm K}\pi$, and the saturation value of their respective integrals exhibit a small, but finite, dependence of the PYTHIA mode used in the simulations. However, note that the shape of $B^{{p}\pi,s}$ exhibits a much stronger dependence on the tune and modes. Indeed, while $B^{{p}\pi,s}$ obtained with the Monash tune and Shoving mode have nearly same shapes and amplitudes (and thus equal integrals), the Ropes mode produces a relatively large increase of the BF's amplitude and its integral at all values of $\Delta y$. The Ropes mode, which is nominally designed to produce a sizable increase in the strange particle production, features a rather small change of $B^{{\rm K}\pi}$ but a sizable increase of $B^{{p}\pi,s}$. It is unclear to these authors whether such a variation is expected. 

Overall, we observe that the charge balance functions of light charged hadrons, computed with PYTHIA8, feature a finite albeit modest dependence of the collision energy and the specific tune and modes used for their computations. This indicates that BFs of light charged hadrons are sensitive to the details of the particle production models used towards their description. It shall be interesting, in future works, to explore how other models differ in their predictions of the shapes and relative integrals of these balance functions and thus the extent to which balance functions can be genuinely used to probe the particle production and constrain existing MC models. In the next paragraph, we investigate whether baryon balance functions can be useful and sensitive probes of baryon production.

\section{Baryon Balance Functions}
\label{sec:Baryon BF} 

The production of non-strange and strange (light) baryons has been measured in pp as well as in A--A collisions at several beam energies but the underlying mechanisms of their production are yet to be fully elucidated. Models aiming to describe the production of baryons are mostly phenomenological in nature and belong essentially into three categories: string fragmentation~\cite{Skands:2014pea,Bellm:2015jjp,Engel:1994vs}, thermal production~\cite{Andronic:2021dkw}, and hydrodynamical+thermal production~\cite{Du:2023gnv,Nijs:2020roc,Putschke:2019yrg}. There also exists hybrid models involving core (hydrodynamical+thermal production) and corona components (Pomeron exchanges)~\cite{Werner:2023jps}. 
It appears that thermal and hydrodynamical models perform better in large systems while the string based description fairs better in smaller systems although EPOS4~\cite{Werner:2023jps} has had great success reproducing data measured in both small and large collision systems. One might then expect, as per the argument initially set forth in Ref.~\cite{Bass:2000az}, that longitudinal baryon balance functions should be quite broad in light collision systems, particularly in pp collisions, and much narrower in A--A collisions, in part because of late baryon antibaryon (B$\rm \bar B$) production, and in part, as a result of the development of large transverse radial flow in mid to central A--A collisions. One must also consider mechanisms of nuclear stopping. What are indeed the mechanisms at play in (partial) baryon stopping and the production of baryons at central rapidities? Some models invoke fluctuations in the fragmentation of strings~\cite{Skands:2014pea,Bellm:2015jjp}, others involve baryon junctions~\cite{Vance:1997th}, while recent speculations suggest the baryon number might be carried by gluons~\cite{Frenklakh:2023pwy}. Although considerable efforts have been invested in measurements of baryons, including single $p_{\rm T}$ spectra, nuclear modification factor, and anisotropic flow~\cite{ALICE:2022wpn}, it remains unclear whether these results can provide sufficiently constraining information to discriminate between the many theoretical approaches and models mentioned above. There is thus plenty of room to consider new and additional ways of measuring baryon production and elucidate the mechanisms behind it. One such new class of measurements is based on the unified balance functions discussed in this work.

We posit that measurements of BFs over a wide range of rapidity and momentum might shed new and additional information to constrain or tune models of baryon production. By construction, BFs are probing the likelihood one particular baryon might be produced in association (and thus correlated) with another baryon. BFs are also sensitive to the manner in which these baryons might be produced. Indeed, are the BFs narrow or broad in relative rapidity $\Delta y$? Are the baryons produced at the onset of collisions or much later as partons combine near freeze-out? Is the baryon number carried by quarks or gluons? While it is unlikely that measurements of BFs can fully elucidate these questions, they might provide useful additional light to inform phenomenological models. It is thus of interest, in this section, to examine what measurements of baryon BFs are possible and how they should be best performed, i.e., what acceptances, data sizes, and instrumental properties are required to successfully achieve these measurements.

In the context of this work, and for the sake of simplicity, we assume that it is possible to measure low-mass non-strange and weakly decaying  baryons (and their respective anti-baryons) based on the (most probable) decay channels listed in Tab.~\ref{tab:baryons}. Clearly, the technical challenges associated with measurements of neutrons and photons are quite considerable. Although high-energy neutrons can be efficiently  detected with hadronic calorimeters~\cite{Armstrong:1998qs}, implementing such techniques with sufficient angular granularity and low $p_{\rm T}$ threshold may forever remain a challenge.  Likewise, detecting high-energy decay photons, event-by-event, with sufficient resolutions and without background is also rather difficult. Measurements of BFs involving some of the baryons listed in Tab.~\ref{tab:baryons} may thus be particularly challenging. We nonetheless include all these baryons (and corresponding anti-baryons) towards computations of mixed pairs BFs and integrals presented in this work. Additionally, simulations are conducted with a full $p_{\rm T}>0$ range (see \cite{ALICE:2019avo} for  $p_{\rm T}$ ranges achieved by the ALICE experiment). This enables an illustration, in particular,  of baryon balance function sum-rules in the context of the production of non-strange and strange low-mass baryons.  Specifically, we show, in the following, that the computation of BFs yields meaningful integrals and sum of integrals that could in principle be exploited to further elucidate baryon production mechanisms. Additionally note that while it might be desirable to carry out a similar exercise for strange balance functions, a full sum rule cannot be formulated for strangeness because the neutral kaon and its anti-particle, $\rm K^0$ and  $\rm \bar{K}^0$ mix into weak eigenstates ${\rm K}_S^0$ and ${\rm K}_L^0$. 

Our analysis of pp collisions produced with PYTHIA8 is accomplished by turning off the decay of the weakly decaying baryons, while short-lived strange and non-strange hadron resonances (e.g., $N^*$, $\Delta^{++}$, etc.) are allowed to  decay (i.e., via the strong interaction) into lighter hadrons. The baryons listed in Tab.~\ref{tab:baryons} are considered endpoints of the decay sequences of heavier baryons and one thus expects the sums of balance functions, with a common reference particle, to add to unity as an explicit manifestation of baryon conservation, as per the sum rule 
\begin{align}
\label{eq:BF-Sum-Rule}
    \sum_{\bar\alpha} \int B^{{\bar \alpha} \beta,s}(\Delta y){\rm d}\Delta y = 1,
\end{align}
where $\beta$ represents a specific baryon (anti-baryon) species, e.g., a proton, and $\sum_{\bar \alpha}$ represents  a sum over all (low mass) anti-baryons (baryons) listed in Tab.~\ref{tab:baryons}.

\begin{table}
    \centering
    \begin{tabular}{cll}
     \hline
Species     &$c\tau$ (m)       &Observation Method \\ 
\hline
p           & long lived       & spectrometer    \\ 
n           & $\tau = 877.8$ s & hadronic calorimeter  \\ 
$\Lambda^0$ & 0.079            & $\Lambda^0 \rightarrow p+\pi^-$ \\ 
$\Sigma^-$  & 0.045            & $\Sigma^- \rightarrow n+\pi^-$  \\ 
$\Sigma^0$  & 0.022 nm         & $\Sigma^0 \rightarrow \Lambda^0+\gamma$  \\ 
$\Sigma^+$  & 0.024            & $\Sigma^+ \rightarrow p+\pi^0$       \\ 
$\Xi^-$     & 0.049            & $\Xi^- \rightarrow \Lambda^0+\pi^-$ \\ 
$\Xi^0$     & 0.087            & $\Xi^0 \rightarrow \Lambda^0+\pi^0$ \\  
$\Omega^-$  & 0.024            & $\Omega^-\rightarrow\Lambda^0+K^-$  \\
\hline
    \end{tabular}
    \caption{Low mass baryons (and their respective anti-particles) used in the computation of mixed balance functions presented in this work. }
    \label{tab:baryons}
\end{table}

\begin{figure}[!ht]
	%	\centering
\includegraphics[width=0.48\linewidth,trim={4mm 1mm 9mm 3mm},clip]
    {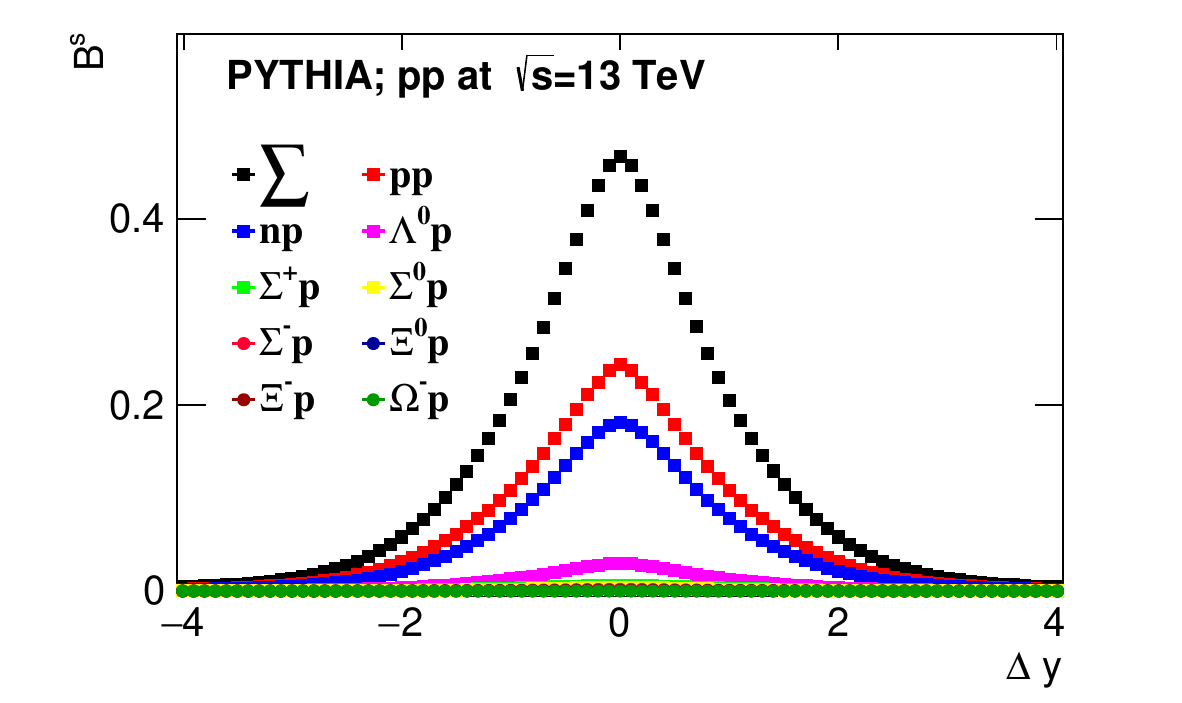}
\includegraphics[width=0.48\linewidth,trim={4mm 1mm 9mm 3mm},clip]
    {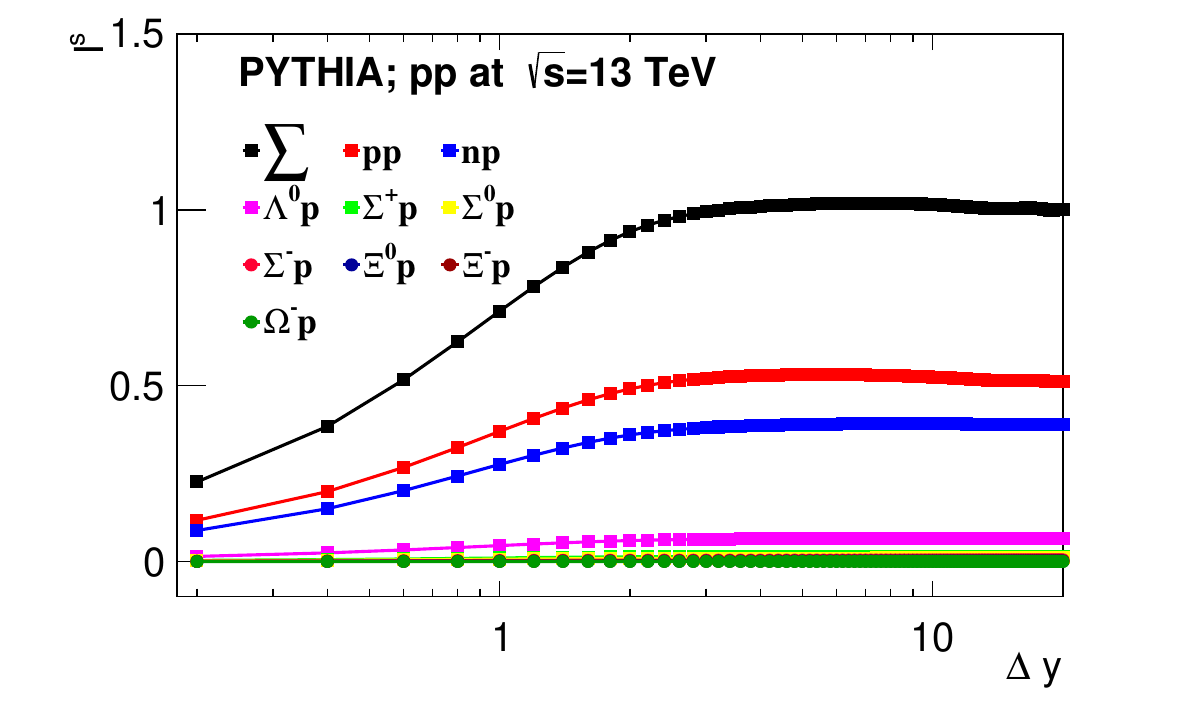}
\includegraphics[width=0.48\linewidth,trim={4mm 1mm 9mm 3mm},clip]
    {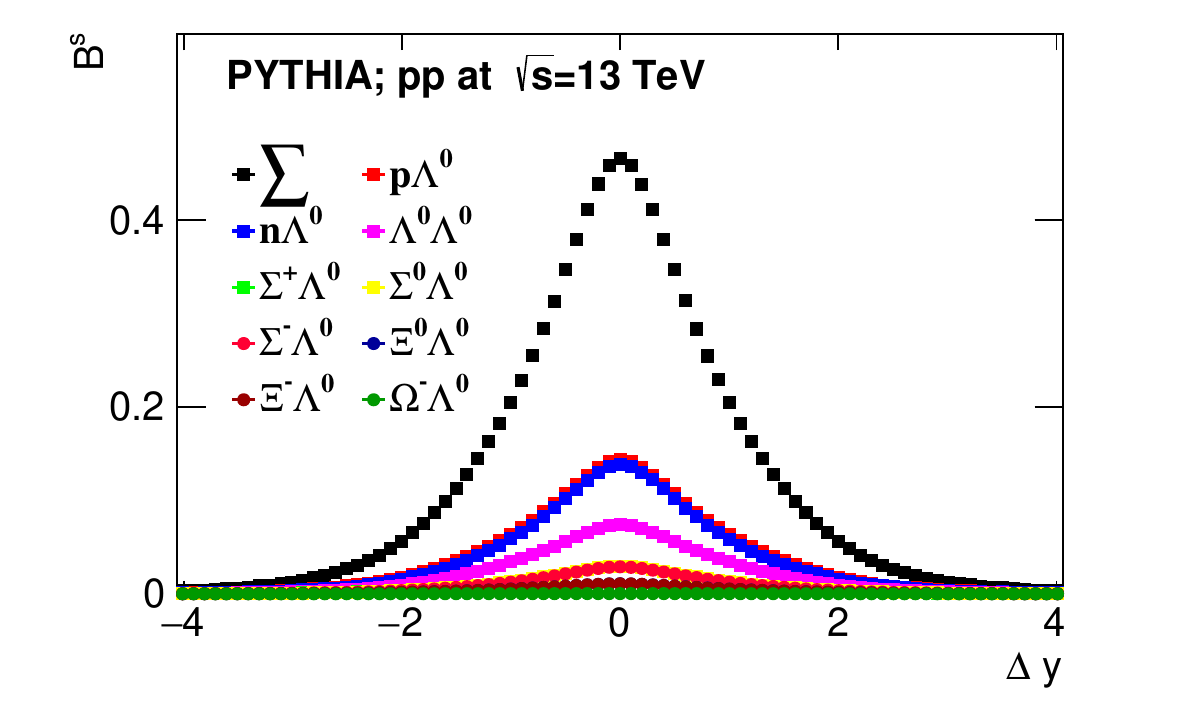}
\includegraphics[width=0.48\linewidth,trim={4mm 1mm 9mm 3mm},clip]
    {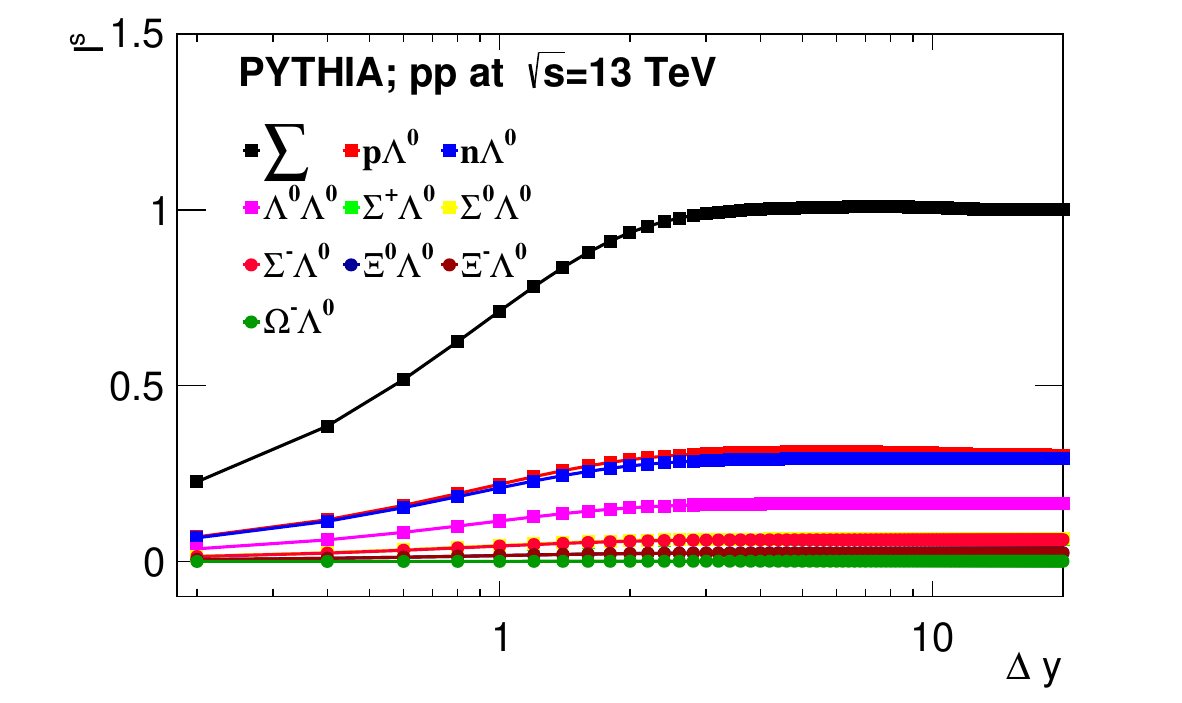}
\includegraphics[width=0.48\linewidth,trim={4mm 1mm 9mm 3mm},clip]
    {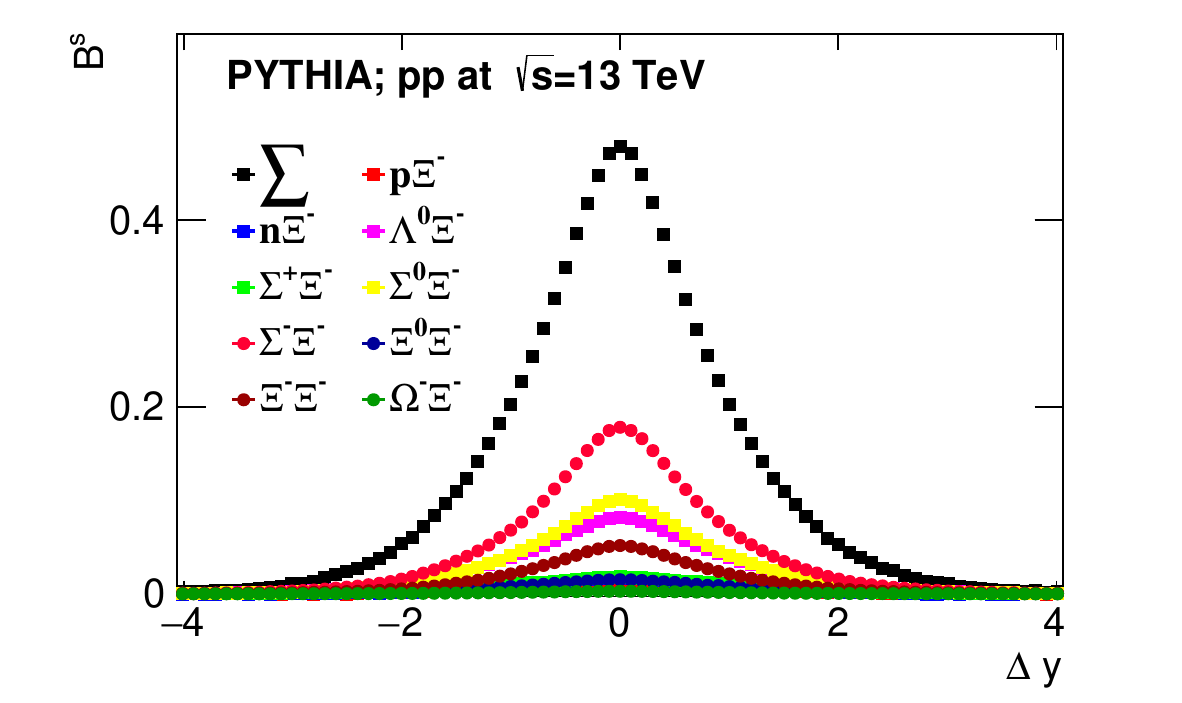}
\includegraphics[width=0.48\linewidth,trim={4mm 1mm 9mm 3mm},clip]
    {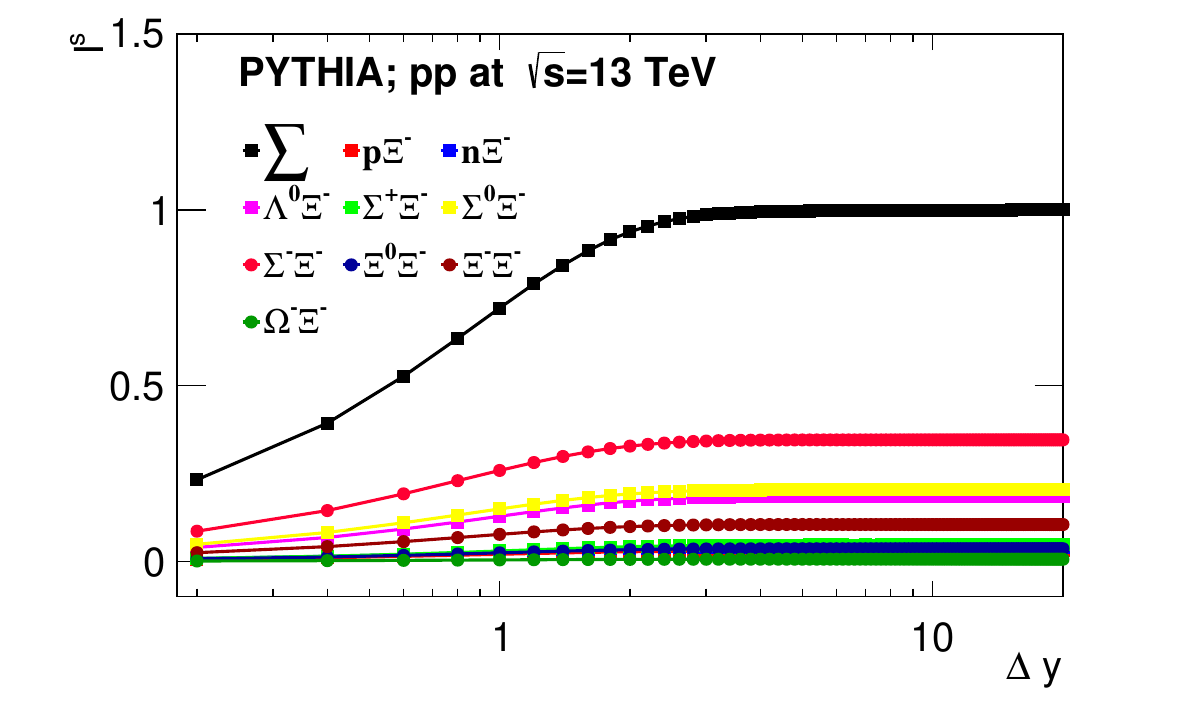}
\includegraphics[width=0.48\linewidth,trim={4mm 1mm 9mm 3mm},clip]
    {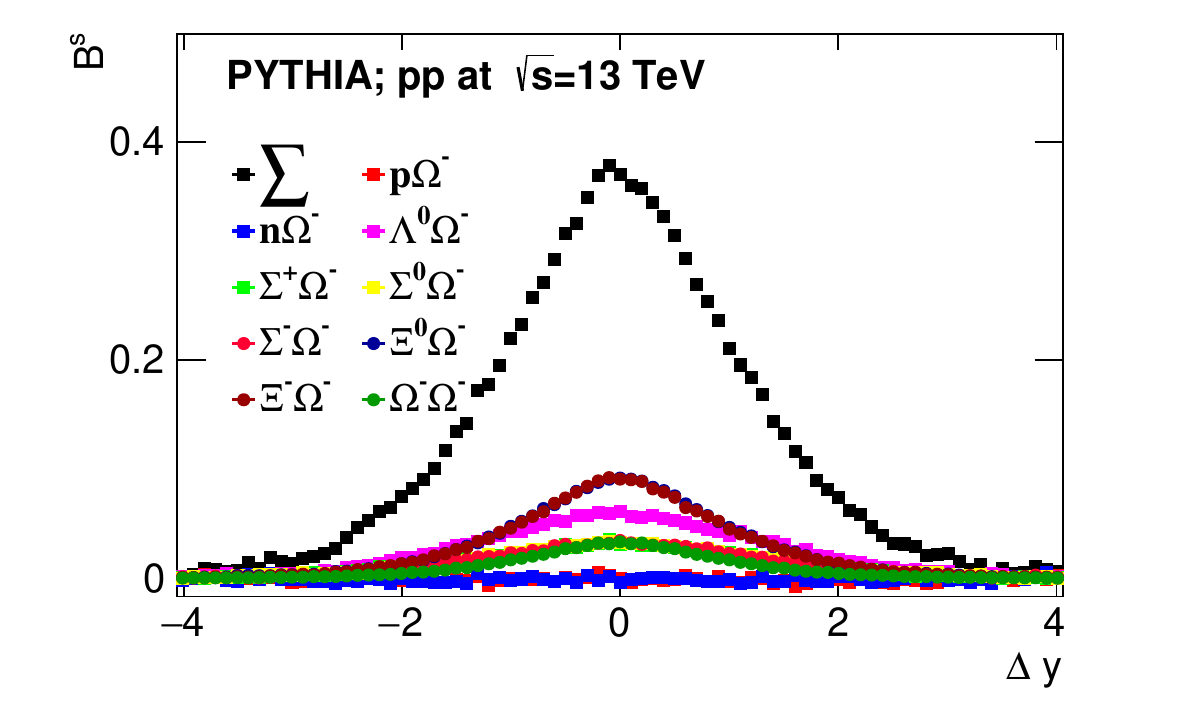}
\includegraphics[width=0.48\linewidth,trim={4mm 1mm 9mm 3mm},clip]
    {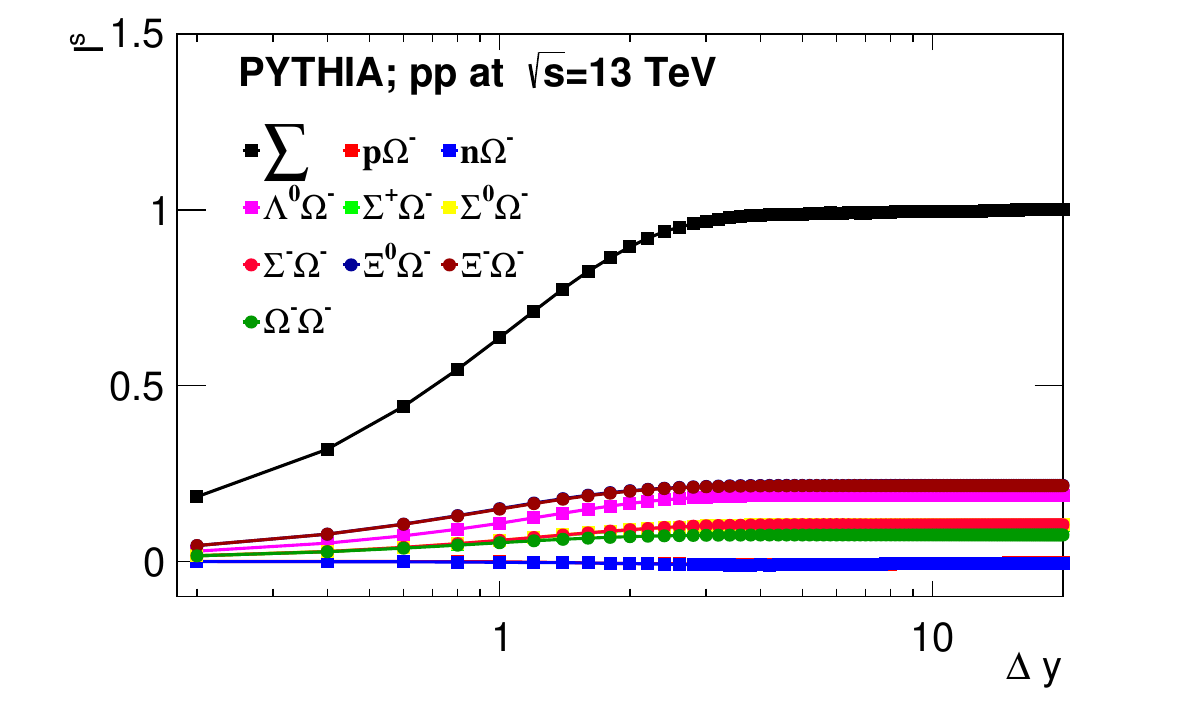}
\caption{Selected mixed pairs baryon balance functions $B^{\alpha\beta,s}$ computed from pp collisions at $\sqrt{s}=13$ TeV simulated with PYTHIA, Monash tune, with color reconnection.}
	\label{fig:baryonsBs}
\end{figure}

Figure~\ref{fig:baryonsBs} presents selected (left) {\bf baryon} balance functions ($B^{\alpha\beta,s}$), and (right) their respective cumulative integrals ($I^{\alpha\beta,s}$) computed with Eqs.~(\ref{eq:Bs},\ref{eq:CumulativeIntegralBF}), for species pairs involving a proton, a $\Lambda^0$, a $\Xi^-$, and an $\Omega^-$ (or their respective anti-particles) as reference. These BFs are computed with PYTHIA8 (Monash tune, with color reconnection) for pp collisions at $\sqrt{s}=13$ TeV. Perfect acceptance (i.e., $p_{\rm T}>0$ and $4\pi$ coverage) is again here assumed in the calculations.  
We find that the amplitudes of the baryon BFs vary greatly with both the reference baryons and their balancing partners. One  notes, in particular, that the baryon number balancing evolves considerably with the reference baryon. For instance, protons ${\rm p}$ are most often balanced by anti-protons $\rm \bar{p}$, but balancing by emission of anti-neutrons ${\rm \bar{n}}$ is a close second, while heavier (strange) baryons contribute much less. 
Interestingly, $\Lambda^0$ ($\bar{\Lambda}^0$), are approximately equally balanced by ${\rm \bar{p}}$ ($\rm p$) and ${\rm \bar{n}}$ ($\rm n$), whereas balancing by $\bar{\Lambda}^0$ ($\Lambda^0$ ) comes in third place. For heavier strange baryons, the balancing scheme changes significantly: balancing by ${\rm \bar{p}}$ ($\rm p$) and ${\rm \bar{n}}$ ($\rm n$) becomes less probable and most of the balancing is achieved with significant probability by one or more anti-strange baryons. Baryon balancing of $\Omega^-$ (${\bar\Omega}^-$) is a somewhat extreme case with essentially no balancing yield from ${\rm \bar{p}}$ ($\rm p$) and ${\rm \bar{n}}$ ($\rm n$). This should not be much of a surprise, however, given the strangeness content of $\Omega^-$ must also be balanced. Evidently, protons and neutrons do not carry strangeness and thus cannot directly balance the strangeness content of $\Omega^-$. One could nonetheless envision more complex balancing schemes where the proton (neutron) balances the baryon number while strangeness balancing is achieved by strange mesons. Such complicated schemes are manifestly not favored by PYTHIA8's reconnection scheme.
We also observe that within PYTHIA8 the most probable mechanism to balance a baryon (anti-baryon)  with strangeness $|S|$ involves a baryon carrying $|\bar S|-1$, independently of whether the reference baryon (anti-baryon) carries one, two or three units of strangeness (anti-strangeness). This constitutes a very specific prediction that should be checked experimentally and would thus provide a strong test of the adequacy of the baryon and flavor production schemes used in PYTHIA8.

Cumulative integrals of the mixed baryon BFs are presented in the right panels of Fig.~\ref{fig:baryonsBs}. One first observes that all cumulative integrals rise rapidly in magnitude up to the acceptance width $\Delta y\sim 2$ and thereafter slowly up to their maximum value in the limit $\Delta y \sim 20$. We also observe that, as for BFs amplitudes, the integral of the various pairs yield a variety of fractional values (i.e., in the range $0<I^s\le 1$) that depend both on the reference and the associated particles. As expected, we also find that the sum of the cumulative integrals, for a specific reference particle,  add to unity in the full acceptance. The baryon balancing sum-rule~(\ref{eq:BF-Sum-Rule}) is indeed satisfied with the selected baryons. 

%Our analysis indicates these BFs are typically non-Gaussian vs. $\Delta y$, owing to ``long" side-tails, but they are reasonably well described by generalized-Gaussian functions of the form 
%\begin{align}
%    B^s(\Delta y) = A \frac{\gamma_{y} \omega_{y}}{2\Gamma(1/\gamma_y)} \exp\left[ -\left( \frac{ \Delta y}{\omega_y} \right)^{\gamma_y}\right],
%\end{align}
%with rms width, $\sigma_y$, calculated according to  
%\begin{align}
%    \sigma_y = \sqrt{\frac{ \omega_y^2 \Gamma(3/\gamma_y)}{\Gamma(1/\gamma_y)}}. 
%\end{align}
%Widths of the BFs are shown in Fig. XXXX. {\color{blue} A bit more discussion needed here]}

\begin{figure}[!ht]
	%	\centering
\includegraphics[width=0.48\linewidth,trim={4mm 1mm 9mm 3mm},clip]
    {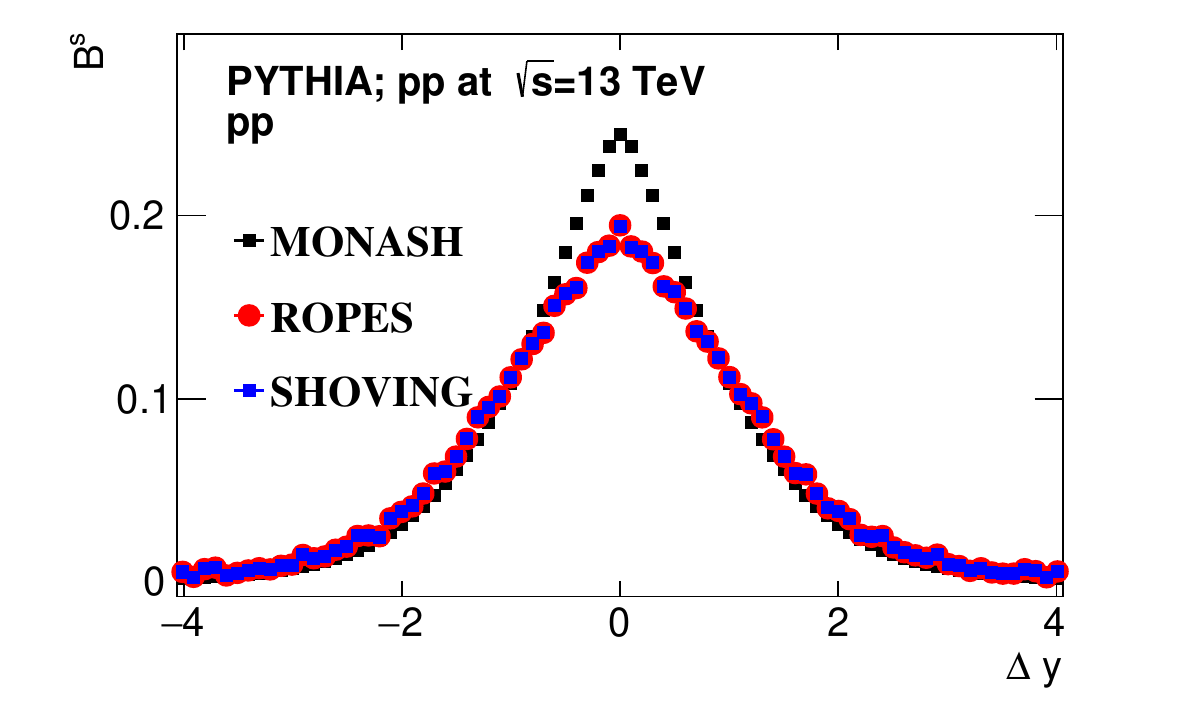}
\caption{Proton-proton baryon balance function $B^{pp,s}$ computed from pp collisions at $\sqrt{s}=13$ TeV simulated with PYTHIA8, using the Monash tunes (Black) and the Ropes (Red), and Shoving (Blue) modes.}
	\label{fig:baryonsBsVsTune}
\end{figure}

We next consider the evolution of baryon BFs with PYTHIA8 tune and modes. Figure~\ref{fig:baryonsBsVsTune} display the  proton-proton  balance functions $B^{{\rm pp},\rm s}$  computed for pp collisions at $\sqrt{s} = 13$  TeV. Other baryon balance functions were found to have a similar dependence of the tune and modes and are thus not shown. The amplitude  of the BFs obtained with the Ropes and Shoving modes are essentially identical   but feature an amplitude at $\Delta y=0$ significantly smaller than that observed with the Monash tune. This suggests that baryon balance functions are sensitive to the color reconnection scheme used to produce baryons but not to the details of Ropes and Shoving modes implementations. Clearly the amplitudes and shapes of baryon balance functions are sensitive to some aspects of the baryon production mechanisms and it will be of interest, in future studies, to explore how other models compare to those of PYTHIA8.
\section{Summary}
\label{sec:summary} 

We presented a Monte Carlo based study of mixed species charge and baryon balance functions (BFs) to explore whether measurements of such BFs could provide useful (new) information on mechanisms of particle production in pp collisions. The study was based on simulations of pp collisions at selected values of $\sqrt{s}$ with PYTHIA8 operated with the Monash tune and the Ropes, and Shoving modes. These simulations showed charge balance functions of mixed pairs of pions, kaons, and protons have distinct shapes and amplitudes that are sensitive to both the collision energy and the model mode. The evolution of their cumulative integral with the acceptance of the measurement is found, in particular, to have good sensitivity to both the beam energy and the details of string fragmentation and color reconnection. Measurements of such correlation functions in pp and larger system thus indeed stand to provide new and useful information to further the  understanding of particle production and transport in small to large collision systems at both RHIC and LHC energies.  We next proceeded to conduct a concept analysis of mixed baryon balance functions. BFs were computed for a selection of light baryon pairs, including neutrons and strange hadrons that might be difficult to measure in practice in the context of similar analyses. We showed that these BFs have amplitudes, shapes, and cumulative integrals that depend on the reference particle as well as the balancing partner. We additionally verified, in the concrete context of PYTHIA8 simulations, that the mixed baryon BFs obey a sum rule constrained by baryon number conservation.    
Finally, we saw that the shape and integrals of the mixed baryon BFs are sensitive to energy and details of particle production. We saw, in particular, that according to PYTHIA, the production of a baryon (anti-baryon) with strangeness $|S|$ is most often balance with an anti-baryon (baryon) with $|\bar S|-1$. We conclude that mixed pairs charge and baryon balance functions might indeed constitute a useful tool in the study of small to large nucleus--nucleus collisions to further the understanding of charged particle production as well as baryon stopping and baryon pair production. 

This study was restricted, for simplicity, to simulations based on three recent modes of PYTHIA8. Given, the mixed species charge and balance functions studied with PYTHIA8 show  sensitivity to the details of the mode as well as the collision energy, one must endeavors to also study their behavior in the context of other models. Such additional studies are in progress and should provide proper background and references for recent and upcoming measurements of mixed species balance functions at RHIC and LHC.

\newenvironment{acknowledgement}{\relax}{\relax}
\begin{acknowledgement}
\section*{Acknowledgements}
This work was supported in part by the United States Department of Energy, Office of Nuclear Physics (DOE NP), United States of America, under grant No. DE-FG02-92ER40713.
SB acknowledges the support of the Swedish Research Council (VR) and the Knut and Alice Wallenberg Foundation. AFD and AM were supported by a grant of the Ministry of Research, Innovation and Digitization, CNCS--UEFISCDI, Romania, project number PN-III-P4-PCE-2021-0390, within PNCDI III.

\end{acknowledgement}

\bibliography{main}

\end{document}